\documentclass[12pt,preprint]{emulateapj}
\usepackage{graphicx,natbib,apjfonts}
\usepackage{amsmath}
\usepackage{amssymb}
\usepackage{epstopdf}
\usepackage{xfrac}

\newcommand{\XSPEC}{{\sc XSpec}}
\newcommand{\RXTE}{\emph{RXTE}}
\newcommand{\erf}{\text{erf}}
\slugcomment{ApJ (accepted)}

\defcitealias{WorpelEtAl2013}{WGP13}
\defcitealias{GallowayEtAl2008}{G08}

\begin{document}

\date{\today}
\title{Evidence for enhanced persistent emission during sub-Eddington thermonuclear bursts}
\author{
Hauke Worpel\altaffilmark{1,2},
Duncan K. Galloway\altaffilmark{1},
Daniel J. Price\altaffilmark{1}
}
\altaffiltext{1}{School of Physics and Astronomy, Monash University, Clayton, Victoria 3800, Australia}
\altaffiltext{2}{Leibniz-Institut f\"ur Astrophysik Potsdam (AIP), An der Sternwarte 16, 14482 Potsdam, Germany}
\begin{abstract}
In a recent paper, we found evidence for an increase in the accretion rate during photospheric radius expansion bursts, quantified by a variable normalization factor $f_a$ on the pre-burst persistent emission. Here we follow this result up on a much larger sample of 1759 type I X-ray bursts from 56 sources. We show that the variable persistent flux method provides improvements in the quality of spectral fits for type I bursts, whether or not they reach the Eddington luminosity. The new approach has an estimated Bayes factor of 64 improvement over the standard method, and we recommend the procedure be adopted as standard for analysing type I bursts. We show evidence that the remaining discrepancies to a formally consistent spectral model are due to the burst component deviating significantly from a blackbody, rather than variations in the spectral shape of the persistent emission component. In bursts that do not show radius expansion, the persistent emission enhancement does not exceed 37\% of the Eddington flux. We use this observation to constrain the Eddington flux of sources for which $F_\text{Edd}$ has not been directly measured.
\end{abstract}
\section{Introduction}
\label{sec:intro}

Type I X-ray bursts, discovered in the 1970s \citep{GrindlayEtAl1976, BelianEtAl1976}, arise from thermonuclear runaways in accreted H/He material on the surface of a neutron star (e.g., \citealt{WoosleyTaam1976,StrohmayerBildsten2006}). Gas flowing from a low-mass stellar companion accumulates on the surface of the neutron star (e.g., \citealt{WoosleyTaam1976,Joss1977}) . There it undergoes hydrostatic heating and compression, triggering unstable nuclear burning when the temperature and pressure are high enough (e.g., \citealt{FujimotoEtAl1981, StrohmayerBildsten2006}). These events are observed as a sudden increase in the neutron star's X-ray luminosity, to many times the persistent level (see reviews by \citealt{LewinJoss1981,LewinEtAl1993}). Typical bursts exhibit rise times of 1--10~s, durations of a few tens of seconds to a few minutes (e.g., \citealt{GallowayEtAl2008}), and emit a total energy of $10^{39}$--$10^{40}$~erg.

Type I bursts provide a probe into the conditions on the surface of the neutron star and yield insights into its internal structure \citep{LattimerPrakash2007}. Performing such measurements requires an accurate determination of the flux throughout the burst. This in turn necessitates a good understanding of the X-ray spectra since bolometric flux cannot be measured directly from photon count \citep{BlissettCruise1979}, and we must also know how to separate the flux from the expanding atmosphere from that arising from other sites in the neutron star system. One such contribution is the accretion disc itself. It emits X-rays through the conversion of gravitational potential energy of the inspiralling gas. This emission is present before the burst, and has traditionally been assumed to remain constant (both in spectral shape and in intensity) throughout the burst (e.g., \citealt{vanParadijsLewin1986,LewinEtAl1993,KuulkersEtAl2003,GallowayEtAl2008}).

\cite{MunoEtAl2000} and \cite{StrohmayerBrown2002} allowed for the persistent emission to be suppressed entirely, by subtracting the instrumental background only, but did not find that the spectral fits were improved by doing so. In a study of neutron star radii derived from cooling tail spectra, \cite{GuverEtAl2012} noted the possibility of variations in the persistent emission spectrum and excluded bursts whose pre-burst persistent flux was more than 10\% of the Eddington flux of that source, so that the contribution from persistent emission is always small. \cite{intZandEtAl2013} studied a PRE burst from SAX~J1808.4$-$3658 using combined \emph{Chandra} and \emph{RXTE} data and found that an observed excess of photons at both low and high energies can be well described by allowing a 20-fold increase of the pre-burst persistent emission. This approach was applied to 332 PRE bursts observed with \emph{RXTE} by \citeauthor{WorpelEtAl2013} (2013; hereafter WGP13), where we found that allowing the persistent emission to vary in intensity during a burst improves the spectral fits, and that the persistent emission usually increases temporarily to several times its pre-burst level. 

Since the intensity of the persistent emission is expected to increase with increasing accretion rate, \citetalias{WorpelEtAl2013} interpreted this effect as a temporary accretion enhancement, possibly the result of the disk losing angular momentum via radiation drag. Other interpretations are possible; \cite{intZandEtAl2013} argue for reprocessing of the burst spectrum in the accretion disc. 

The "variable persistent flux" approach has since been applied successfully by other studies (e.g., \citealt{intZandEtAl2013, KeekEtAl2014, PeilleEtAl2014}), and it is becoming clearer that bursts affect the accretion disc and can be used to probe accretion physics. In an investigation of a long duration burst from 4U~1636$-$536, \cite{KeekEtAl2014} found that the persistent emission was enhanced by a factor of $\approx 2$ and also inferred a significant modification of the inner accretion disc. A depletion of the accretion disc through radiation drag during a burst was predicted by \cite{Walker1992}, who found in numerical simulations that the accretion rate after the burst can decline temporarily to half the pre-burst level. \cite{PeilleEtAl2014} performed a study of quasi-periodic oscillations from 4U~1636$-$536 and 4U~1608$-$522, and found that these oscillations were suppressed for several tens of seconds after the burst. They attribute this phenomenon to a depletion of the inner accretion disc, though its fast recovery (5-10 times faster than the viscous time) remains a mystery. Clearly a greater observational and theoretical understanding of the behaviour of accretion onto type I bursters is required.

Disentangling the burst component of the spectrum from the persistent component is difficult, as the various components are usually spectrally degenerate. Furthermore, we do not have spatial resolution and therefore cannot attribute different spectral components to different sites in the neutron star system. This problem will be further complicated if, as we expect, both the persistent and burst components can vary in spectral shape during a burst. Indeed, it has long been known that the spectral shape of the persistent emission, as well as its intensity, is a function of accretion rate (\citealt{HasingervanderKlis1989}; see also \citealt{LyuEtAl2014}). There are, however, suggestions that between bursts the accretion spectrum remains constant in shape on timescales longer than those of a burst, though its intensity may vary (e.g., \citealt{ThompsonEtAl2005, BagnoliEtAl2013, LinaresEtAl2014}). During a burst it is, of course, much more difficult to detect changes in the accretion spectral shape. A sudden loss of high energy ($\gtrsim 30$~keV) photons has been detected from several sources, but this effect has been attributed to rapid cooling of the corona and is not related to accretion rate \citep{MaccaroneCoppi2003, JiEtAl2013, ChenEtAl2013}. Changes in the shape and intensity of the persistent emission during a very long (20,000s) burst from 4U~1636$-$536 were reported by \cite{KeekEtAl2014}, but variations on these timescales can happen in the absence of bursts, and the bursts we consider in this paper are very much shorter.

PRE bursts change the structure of the neutron star photosphere, and this is likely to cause the spectrum of the photosphere to deviate from its usual near-blackbody shape. Such changes are unrelated to the accretion spectrum. As pointed out by \citetalias{WorpelEtAl2013}, these will confound spectral analyses of persistent emission during a burst. Such complications can be removed by considering non-PRE bursts, but care must be taken to account for lower sensitivity due to their intrinsically lower flux.

In this paper we investigate the relationship between accretion and type I bursts. We apply the fitting method of \citetalias{WorpelEtAl2013} to every type I burst detected by \RXTE. In addition, we investigate the variability of the accretion spectra immediately before and after bursts, to determine whether our spectral fitting procedure can be confounded by rapid changes in the accretion spectrum. In \S \ref{sec:Edd_from_nonpre} we develop a method for obtaining a lower bound on the Eddington flux for sources that have not yet been observed to undergo PRE bursts.

\section{Data analysis}
\label{sec:data}

We used observational data from the \emph{Rossi X-Ray Timing Explorer} (\emph{RXTE}), publicly available through the High-Energy Astrophysics Science Archive Research Centre (HEASARC)\footnote{See http://heasarc.gsfc.nasa.gov}. The observations date from shortly after the satellite's launch on December 30, 1995 to the end of the \emph{RXTE} mission on January 3, 2012. Our sample of type I bursts is based on the burst catalogue of (\citealt{GallowayEtAl2008}; hereafter G08), with the addition of 880 type I bursts detected after the publication of that paper. The entire \RXTE\ burst catalog forms part of the Multi-INstrument Burst Archive (MINBAR)\footnote{ see http://burst.sci.monash.edu/minbar}.

We classified type I bursts as either belonging to the photospheric radius expansion (PRE) class of bursts, or not, according to the criteria for radius expansion described in \S 2.3 of \citetalias{GallowayEtAl2008}. In brief, these criteria define radius expansion as an increase of the surface area of the photosphere together with a decrease in its temperature. A small number of bursts fulfil some, but not all, of the criteria and are classified as "marginal" PRE bursts.

Our data reduction procedures are the same as those in \S 2 of \citetalias{WorpelEtAl2013}, unless otherwise stated, and we refer the reader to that work for further details. To estimate and remove the instrumental background we
used the full-mission, ``bright'' source (>40 counts s$^{-1}$) models released 2006 August 6 with the \texttt{pcabackest} tool. Table \ref{tab:fedd} lists interstellar absorption column densities for all burst sources and the references from which these were drawn; the table includes several sources that were not investigated in \citetalias{WorpelEtAl2013} because they have no PRE bursts in the MINBAR catalog.

\begin{deluxetable*}{lllccccc}
\tabletypesize{\scriptsize}
\tablecaption{$N_H$ Values and Eddington Fluxes}
\tablewidth{0pt}
\tablehead{
\colhead{Source} & \colhead{Non-PRE Bursts} & \colhead{PRE bursts} & \colhead{SWFC\tablenotemark{a}} & \colhead{SBC\tablenotemark{b}} & \colhead{$N_H$} & \colhead{Reference}& \colhead{$F_\text{Edd}$}\\
                  &                           &                       &                                   &                                  &   \colhead{$10^{22}$ cm$^{-2}$}  & & \colhead{$10^{-9}$ erg s $^{-1}$ cm$^{-2}$}}
\startdata
4U 1746-37 & 22 & 3 & 0 & 2 & 0.3 & 1&5.8 $\pm$ 1.4\\
MXB 1659-298 & 11 & 12 & 0 & 0 & 0.2 & 2&17.0 $\pm$ 4.0\\
Aql X-1 & 52 & 12 & 0 & 0 & 0.4 & 3&99.6 $\pm$ 21.3\\
SAX J1810.8-2609 & 4 & 1 & 1 & 0 & 0.3 & 4&111.2 $\pm$ 2.5\\
IGR J17480-2446 & 277 & 0 & 0 & 0 & 0.5 & 1&\nodata\tablenotemark{c}\\
Cyg X-2 & 34 & 7 & 0 & 0 & 0.1 & 5&13.1 $\pm$ 2.1\\
GS 1826-24 & 67 & 0 & 0 & 0 & 0.4 & 6&\nodata\\
XTE J1709-267 & 2 & 0 & 0 & 0 & 0.4 & 7&\nodata\\
4U 1820-30 & 0 & 16 & 34 & 7 & 0.2 & 1&57.1 $\pm$ 7.7\\
SAX 1808.4-3658 & 1 & 8 & 2 & 0 & 0.1 & 8&230.1 $\pm$ 13.2\\
GX 17+2 & 5 & 2 & 0 & 4 & 1.9 & 9&13.3 $\pm$ 2.5\\
Rapid Burster & 19 & 0 & 0 & 0 & 1.7 & 10&\nodata\\
IGR J17473-2721 & 2 & 0 & 0 & 0 & 3.8 & 11&113.5 $\pm$ 12.1\\
Cir X-1 & 13 & 0 & 0 & 0 & 0.7 & 12&\nodata\\
HETE J1900.1-2455 & 0 & 7 & 0 & 0 & 0.2 & 13&123.9 $\pm$ 10.6\\
4U 2129+12 & 5 & 1 & 1 & 1 & 0.0 & 1&40.8 $\pm$ 1.6\\
1M 0836-425 & 17 & 0 & 0 & 0 & 2.2 & 14&\nodata\\
4U 1735-444 & 7 & 10 & 0 & 0 & 0.1 & 15&34.2 $\pm$ 5.6\\
1A 1744-361 & 3 & 0 & 0 & 0 & 4.5 & 16&\nodata\\
4U 1722-30 & 0 & 3 & 23 & 1 & 0.8 & 1&61.7 $\pm$ 12.4\\
GX 3+1 & 1 & 2 & 0 & 1 & 1.6 & 17&59.9 $\pm$ 0.9\\
EXO 1745-248 & 15 & 2 & 0 & 0 & 3.8 & 1&69.0 $\pm$ 0.2\\
XTE J1739-285 & 6 & 0 & 0 & 0 & 2.0 & 18&\nodata\\
2S 0918-549 & 1 & 2 & 3 & 0 & 0.3 & 19&119.2 $\pm$ 12.4\\
Ser X-1 & 11 & 6 & 0 & 0 & 0.4 & 5&29.4 $\pm$ 7.1\\
GRS 1741.9-2853 & 1 & 4 & 1 & 0 & 11.3 & 20&35.3 $\pm$ 10.9\\
2E 1742.9-2929 & 65 & 1 & 0 & 0 & 1.2 & 18&37.8 $\pm$ 1.4\\
4U 1728-34 & 30 & 64 & 0 & 0 & 2.6 & 21&95.0 $\pm$ 8.4\\
XB 1832-330 & 0 & 1 & 0 & 0 & 0.1 & 1&33.7 $\pm$ 4.4\\
IGR J17511-3057 & 7 & 0 & 0 & 0 & 0.6 & 22&\nodata\\
4U 1608-522 & 28 & 18 & 5 & 0 & 0.9 & 23&167.2 $\pm$ 26.0\\
GRS 1747-312 & 1 & 1 & 0 & 0 & 1.4 & 1&13.4 $\pm$ 4.4\\
XTE 1814-338 & 27 & 0 & 0 & 0 & 0.2 & 18&\nodata\\
XTE J1723-376 & 2 & 0 & 0 & 0 & 7.9 & 18&\nodata\\
4U 1254-69 & 3 & 0 & 0 & 0 & 0.3 & 24&\nodata\\
4U 1702-429 & 43 & 5 & 3 & 0 & 1.9 & 18&87.7 $\pm$ 4.5\\
SAX J1750.8-2900 & 3 & 2 & 1 & 0 & 0.9 & 18&54.1 $\pm$ 2.1\\
XTE J1701-462 & 1 & 2 & 0 & 0 & 2.0 & 25&43.4 $\pm$ 1.4\\
XTE J1810-189 & 3 & 1 & 0 & 0 & 4.2 & 26&54.2 $\pm$ 1.8\\
EXO 0748-676 & 143 & 5 & 0 & 0 & 0.8 & 27&46.5 $\pm$ 4.6\\
4U 1916-053 & 1 & 12 & 0 & 0 & 0.3 & 28&30.6 $\pm$ 3.6\\
XTE 2123-058 & 3 & 0 & 0 & 0 & 0.1 & 29&\nodata\\
XTE 1759-220 & 5 & 3 & 0 & 0 & 2.8 & 18&15.7 $\pm$ 0.8\\
4U 1636-536 & 269 & 76 & 2 & 0 & 0.2 & 5&72.6 $\pm$ 9.1\\
IGR 17191-2821 & 5 & 0 & 0 & 0 & 0.3 & 30&\nodata\\
SLX 1744-300 & 10 & 0 & 3 & 0 & 4.5 & 31&13.9 $\pm$ 3.1\\
SAX J1748.9-2021 & 16 & 11 & 0 & 0 & 0.5 & 1&38.0 $\pm$ 6.0\\
4U 1323-62 & 35 & 0 & 0 & 0 & 2.4 & 32&\nodata\\
KS 1731-260 & 21 & 4 & 3 & 0 & 1.3 & 33&48.6 $\pm$ 5.6\\
XTE J1710-281 & 37 & 3 & 0 & 0 & 0.4 & 34&7.1 $\pm$ 1.5\\
IGR J17498-2921 & 0 & 1 & 0 & 0 & 1.2 & 18&51.6 $\pm$ 1.6\\
SAX J1747.0-2853 & 6 & 10 & 2 & 0 & 8.8 & 35&52.5 $\pm$ 7.1\\
SLX 1735-269 & 1 & 0 & 0 & 3 & 0.1 & 36&49.4 $\pm$ 3.9\\
4U 1705-44 & 77 & 4 & 0 & 0 & 1.9 & 37&41.0 $\pm$ 3.8\\
4U 0513-40 & 11 & 4 & 3 & 0 & 0.0 & 1&14.5 $\pm$ 3.5\\
SAX J1806.5-2215 & 4 & 0 & 0 & 0 & 1.0 & 18&\nodata\\
\enddata
\tablerefs{1. \cite{KuulkersEtAl2003}; 2. \cite{OosterbroekEtAl2001}; 3. \cite{CampanaStella2003}; 4. \cite{NatalucciEtAl2000}; 5. \cite{AsaiEtAl2000}; 6. \cite{intZandEtAl1999}; 7. \cite{JonkerEtAl2003}; 8. \cite{WangEtAl2001}; 9. \cite{FarinelliEtAl2007}; 10. \cite{FrogelEtAl1995}; 11. \cite{AltamiranoEtAl2008}; 12. \cite{IariaEtAl2005}; 13. \cite{Campana2005}; 14. \cite{BelloniEtAl1993}; 15. \cite{AugusteijnEtAl1998}; 16. \cite{GavriilEtAl2012}; 17. \cite{OosterbroekEtAl2001b}; 18. J. in 't Zand 2014, private communication; 19. \cite{JuettEtAl2001}; 20. \cite{SakanoEtAl2002}; 21. \cite{DaiEtAl2006}; 22. \cite{PapittoEtAl2010}; 23. \cite{KeekEtAl2008}; 24. \cite{BoirinParmar2003}; 25. \cite{LinEtAl2009}; 26. \cite{KrimmEtAl2008}; 27. \cite{HomanEtAl2003}; 28. \cite{ChurchEtAl1998}; 29. \cite{HynesEtAl2001}; 30. \cite{Klein-WoltEtAl2007}; 31. \cite{MoriEtAl2005}; 32. \cite{ChurchEtAl2005}; 33. \cite{CackettEtAl2006}; 34. \cite{YounesEtAl2009}; 35. \cite{WernerEtAl2004}; 36. \cite{DavidEtAl1997}; 37. \cite{PirainoEtAl2007}}
\tablenotetext{a}{BeppoSAX Wide-Field Camera}
\tablenotetext{b}{Supplemental Burst Catalog (see table \ref{tab:supplemental_bursts})}
\tablenotetext{c}{No PRE burst has been observed from this source}
\label{tab:fedd}
\end{deluxetable*}

\subsection{Burst selection}
\label{sec:BurstSelection}
We restrict our sample of type I bursts to exclude events that are unsuitable for analysis. We discarded  57 bursts for which no Standard-2 data was available, preventing estimation of the instrumental background.
Some burst sources lie in crowded fields containing other LMXBs within the $\sim1^\circ$ \RXTE\ field of view. If the other source(s) were active at the time of observation, then their persistent emission would be confused with that of the burst source and it would be impossible to separate the persistent emission of the burst source from that of the other source(s). Such bursts need to be excluded from consideration. Our procedure for removing source-confused bursts is the same as \citetalias{WorpelEtAl2013} \S 4.1. A total of  168 bursts were excluded as being source confused. Our catalog also contains 16 type I bursts from the Rapid Burster taken during offset pointings to avoid confusion with the nearby 4U~1728$-$34.
We also excluded  35 bursts whose radius expansion status was unclassifiable due to data gaps or very low flux, and 58 bursts classified as "marginal" (i.e. satisfying only some of the criteria of \S 2.3 in \citetalias{GallowayEtAl2008}).
\subsection{Characterizing the persistent emission}
\label{sec:CharPers}
As in the previous studies \citetalias{GallowayEtAl2008} and \citetalias{WorpelEtAl2013} we adopted the integrated X-ray flux for a 16-second interval prior to the start of each burst as the persistent emission. This spectrum includes a time-dependent contribution estimated as in \citetalias{WorpelEtAl2013}. Subsequent model fits to each persistent (and burst) spectrum used the corresponding model spectrum estimated for that burst as background.

\begin{figure}
\includegraphics[width=96mm]{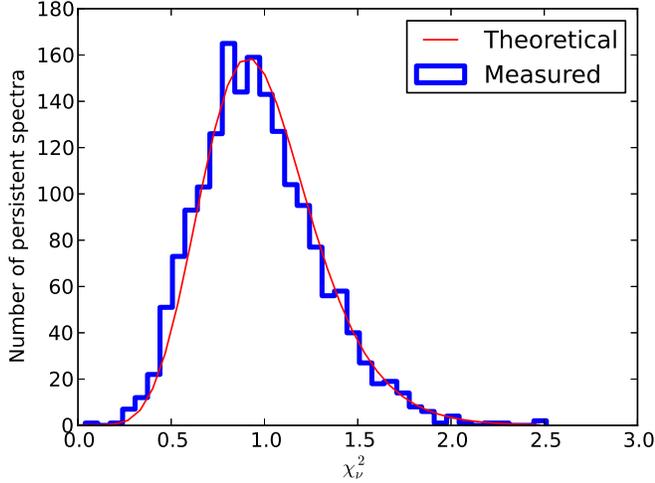}
\caption{Distribution of $\chi^2_\nu$ for the spectral fits to the persistent emission (histogram) compared to a theoretical distribution of $\chi^2_\nu$ for a collection of spectra with the same number of degrees of freedom (curve). A Kolmogorov-Smirnov test gave a 97\% probability ($D=0.11$)that the two curves are consistent with having been drawn from the same distribution, indicating that our suite of models for the persistent emission spectra is adequate for use in subsequent work.}
\label{fig:perschi2}
\end{figure}

For each burst we fit the persistent emission with a set of nine alternative models in turn. These are summarized in Table \ref{tab:Persistent-models}, and are the same six persistent models as in \citetalias{WorpelEtAl2013}, with the addition of \texttt{wabs*disko}, \texttt{wabs*diskm}, and \texttt{wabs*(CompTT+gaussian)}. The first two of these include accretion rate explicitly as a variable parameter, making it possible to investigate the relationship between the normalization factor $f_a$ and $\dot{M}$. The latter was used by \citet{PeilleEtAl2014} and we include it for completeness. We then selected the fit that gave the best (i.e. lowest) $\chi^2_\nu$ to represent the persistent spectrum for that burst. As shown in Figure \ref{fig:perschi2}, this suite of models provide statistically adequate fits so we regard these persistent models acceptable in the analysis of the bursts themselves.

There is no difference in the distributions of best-fitting persistent models between the PRE and non-PRE bursts. We assigned each model a distinct numerical label and performed a Kolmogorov-Smirnov test on the resulting distributions, with $D=0.03$ and a 95\% probability that both are consistent with having been drawn from the same distribution.

A gas pressure dominated accretion disc (\texttt{wabs*diskm}) is not the preferred model for any persistent emission spectrum, indicating that this is not a good description of accretion discs in LMXB systems. Instead, we expect the disc to be radiation dominated, and possibly showing hard emission.

\begin{deluxetable*}{lcl}
\tablecolumns{2}
\tablecaption{XSPEC models for fitting the persistent emission}
\tablehead{
   XSPEC model&
\colhead{Number} &
Notes \\
  & 
  \colhead{of spectra} &
}  
\startdata               
\texttt{ wabs*disko\tablenotemark{a} } & 160 &  \\
\texttt{ wabs*diskm\tablenotemark{a} } & 0 &  \\
\texttt{ wabs*(bbodyrad+powerlaw) } & 425 & nH fixed to literature values \\
\texttt{ wabs*(bbodyrad+powerlaw+gaussian) } & 271 & nH fixed, Gaussian energy set to 6.4keV \\
\texttt{ wabs*(compTT\tablenotemark{b}) } & 39 & nH allowed to vary \\
\texttt{ wabs*(bbodyrad+powerlaw) } & 46 & nH allowed to vary \\
\texttt{ wabs*(gauss+bremss) } & 321 & all parameters variable\tablenotemark{c} \\
\texttt{ wabs*(bbodyrad+diskbb\tablenotemark{d}) } & 183 & all parameters variable\tablenotemark{c} \\
\texttt{ wabs*(compTT+gaussian)\tablenotemark{e} } & 314 & nH fixed, Gaussian energy set to 6.4keV \\
\hline
Total usable spectra & 1759 & \\
\hline
Rejected due to source confusion & 168 & Other active sources in field\\
Background data missing or unusable & 57 & \\
No good persistent model fit & 6 & Minimum $\chi^2_\nu > 3.5$\\
Marginal radius expansion & 58 & \\
Radius expansion status unclassified & 35 & \\
\enddata
\tablenotetext{a}{See \cite{StellaRosner1984}}
\tablenotetext{b}{See \cite{Titarchuk1994b}}
\tablenotetext{c}{All parameters are variable for the generation of persistent emission models. Their values are subsequently frozen for the burst spectral fits in \S \ref{sec:varfafits}}
\tablenotetext{d}{See \XSPEC\ manual (\url{http://heasarc.gsfc.nasa.gov/xanadu/xspec/manual/manual.html}) and references therein}
\tablenotetext{e}{See \cite{PeilleEtAl2014}}
\label{tab:Persistent-models}
\end{deluxetable*}

For  6 bursts, no persistent emission model could be fit with $\chi^2_\nu<3.5$. These bursts were excluded from further consideration.
\subsection{ Fitting procedure}
\label{sec:varfafits}
As in \citetalias{WorpelEtAl2013} we first fit every burst spectrum with the standard approach: a blackbody spectrum, corrected for interstellar absorption using the appropriate hydrogen column density for the burst source (see Table \ref{tab:fedd}). We use the recorded pre-burst emission, including the instrumental background, as the background for these initial fits. See \citetalias{GallowayEtAl2008} for further details regarding this fitting method. The standard fits provide a comparison model and spectral parameters, as well as sensible spectral parameters to seed the variable persistent emission fits.

We then re-fit all the burst spectra, replacing the background with the instrumental background, and adding a variable persistent emission component appropriate for that burst (\S \ref{sec:CharPers}). That is, we modelled the total spectrum as
\begin{equation}
S(E)=A(E)\times B(E;T_\text{bb},K_\text{bb}) + f_a\times P(E)+b(E)_\text{inst},\end{equation} where $S(E)$ is the fitted spectrum as a function of energy $E$, $A$ is the absorption correction due to interstellar hydrogen (see $n_H$ values listed in Table \ref{tab:fedd}), $B$ is a blackbody spectrum with temperature $T_\text{bb}$ and normalization $K_\text{bb}$, and $b_\text{inst}$ is the instrumental background. $P$ is a model for the pre-burst persistent emission that also includes absorption, though for some persistent models we do not retain the same $n_H$-- this is a means to adjust the low energy end of the accretion spectrum and should not be interpreted as saying anything physical about the hydrogen column density. The accretion enhancement is quantified with the $f_a$ normalization factor.
We have fit a total of 160,017 spectra.

\section{Results}
\label{sec:results}

\begin{figure*}
\includegraphics[width=60mm]{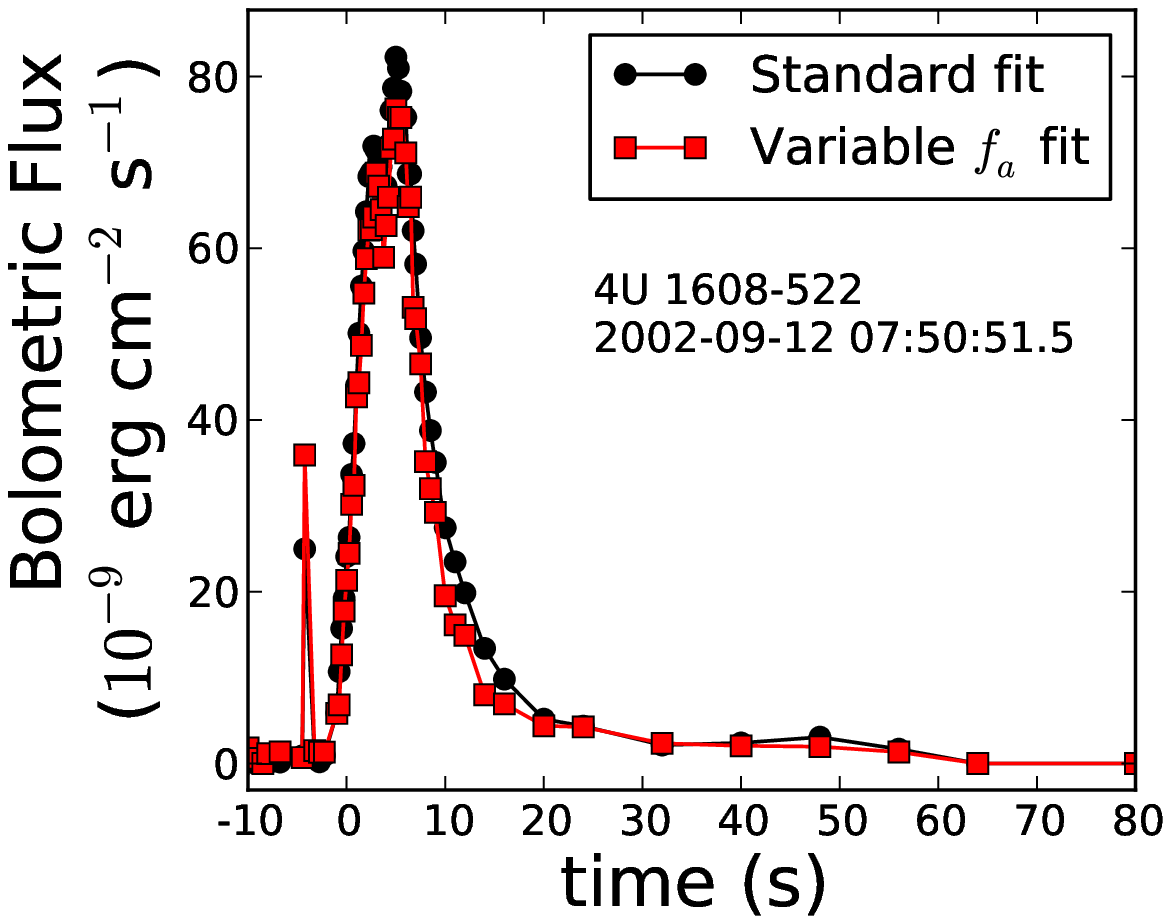}\includegraphics[width=60mm]{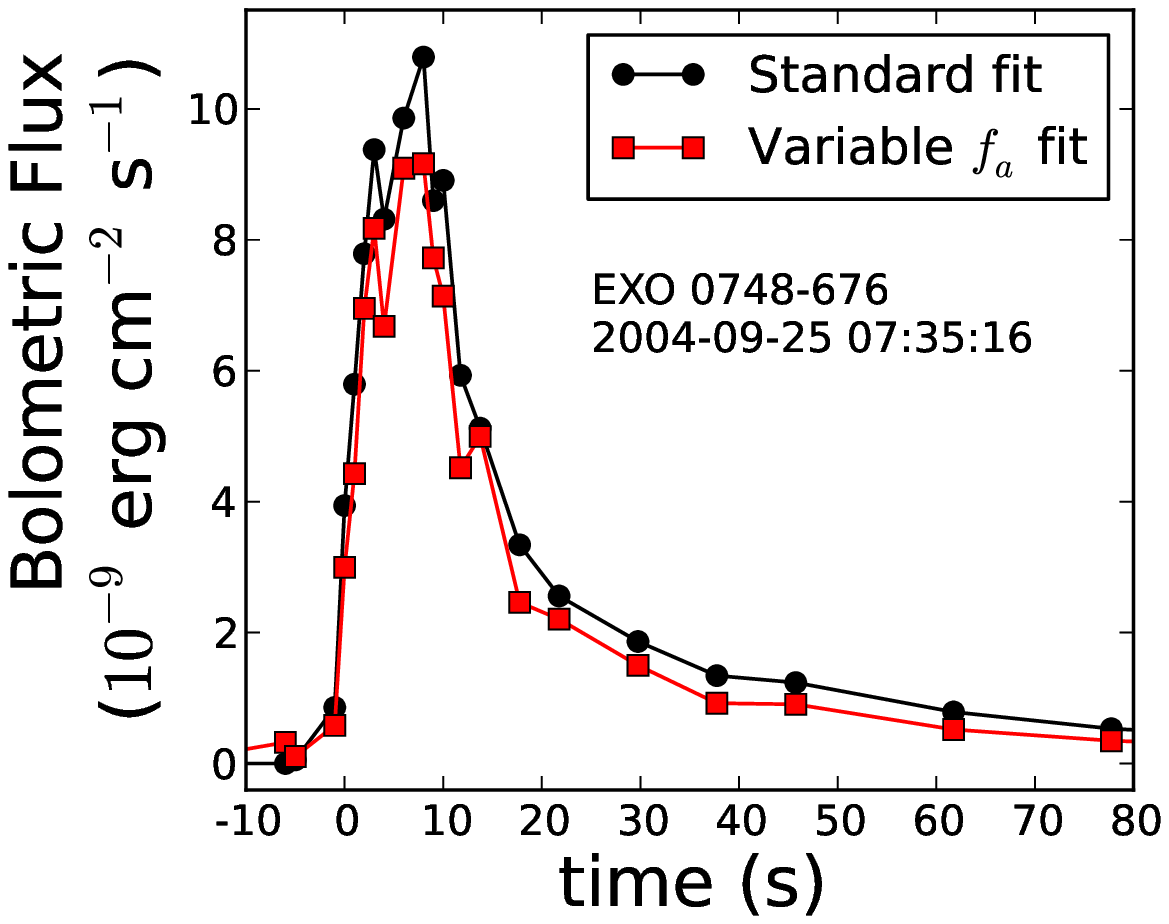}\includegraphics[width=60mm]{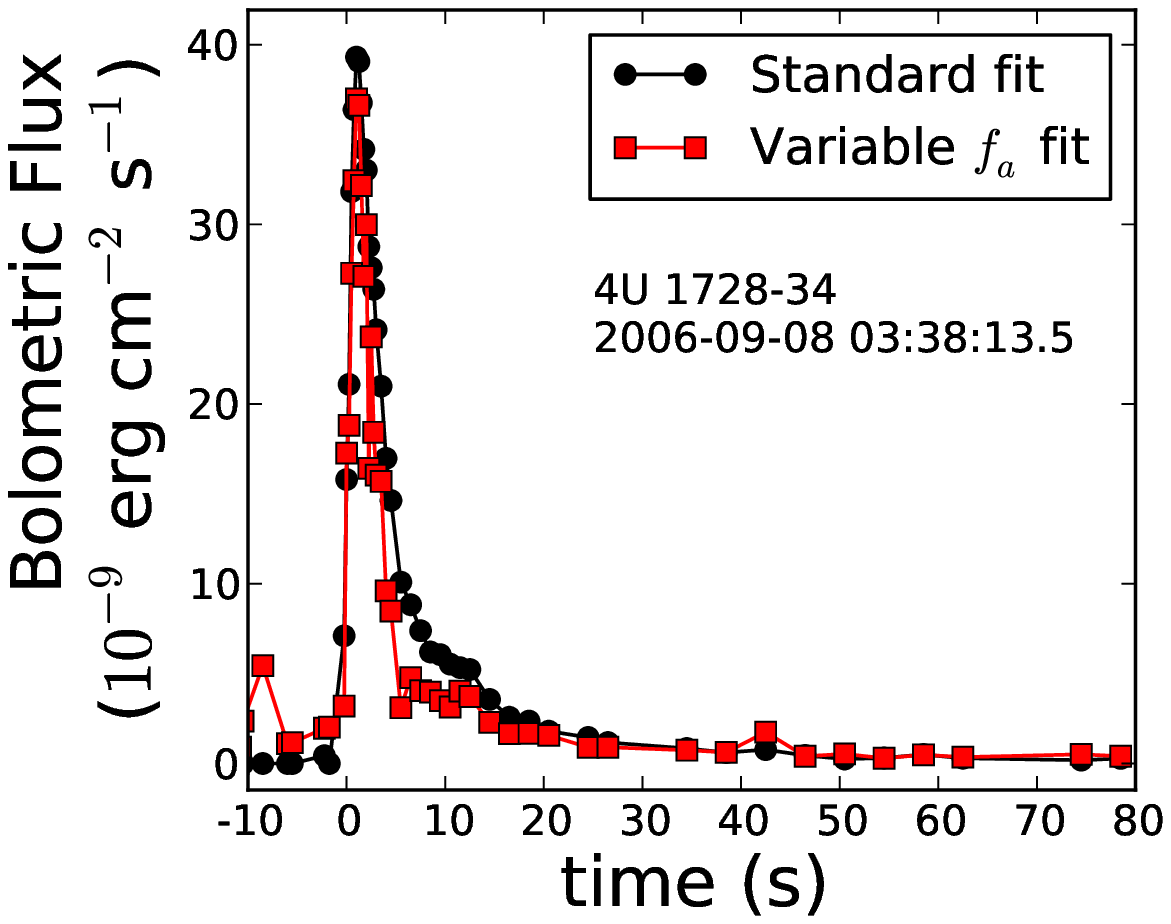}\\
\includegraphics[width=60mm]{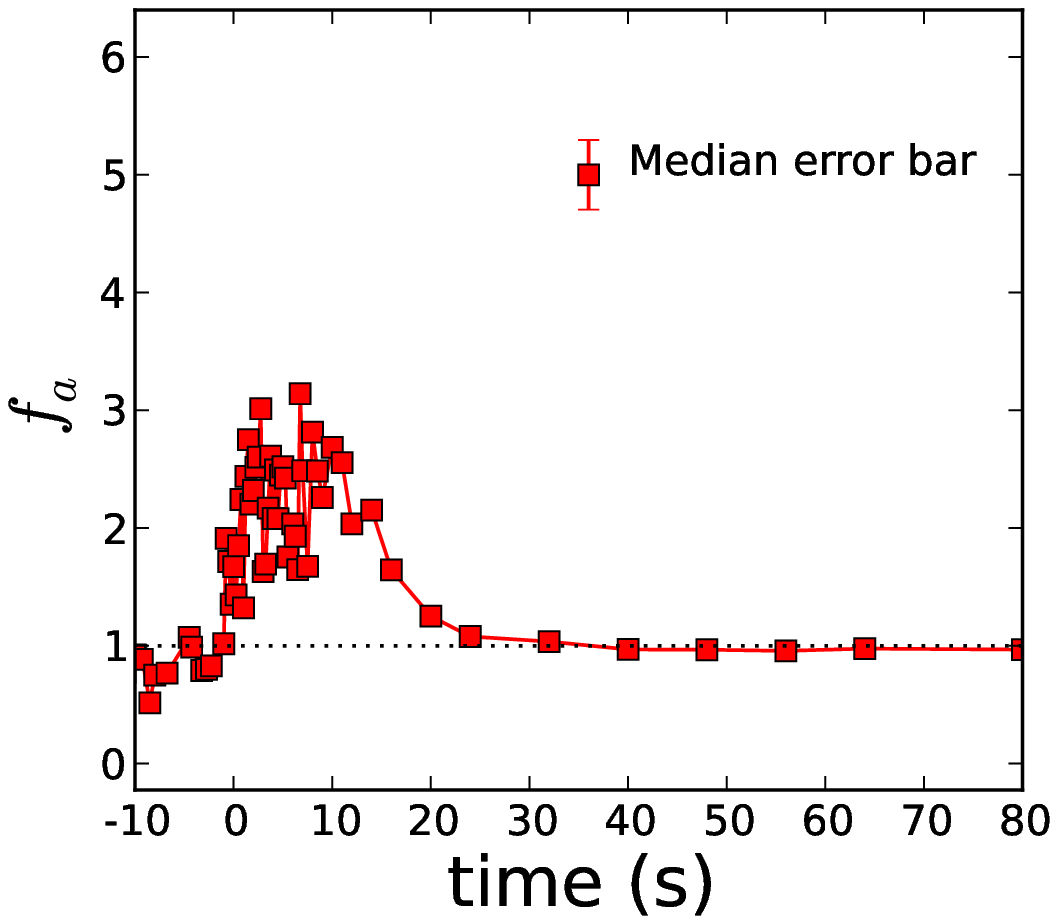}\includegraphics[width=60mm]{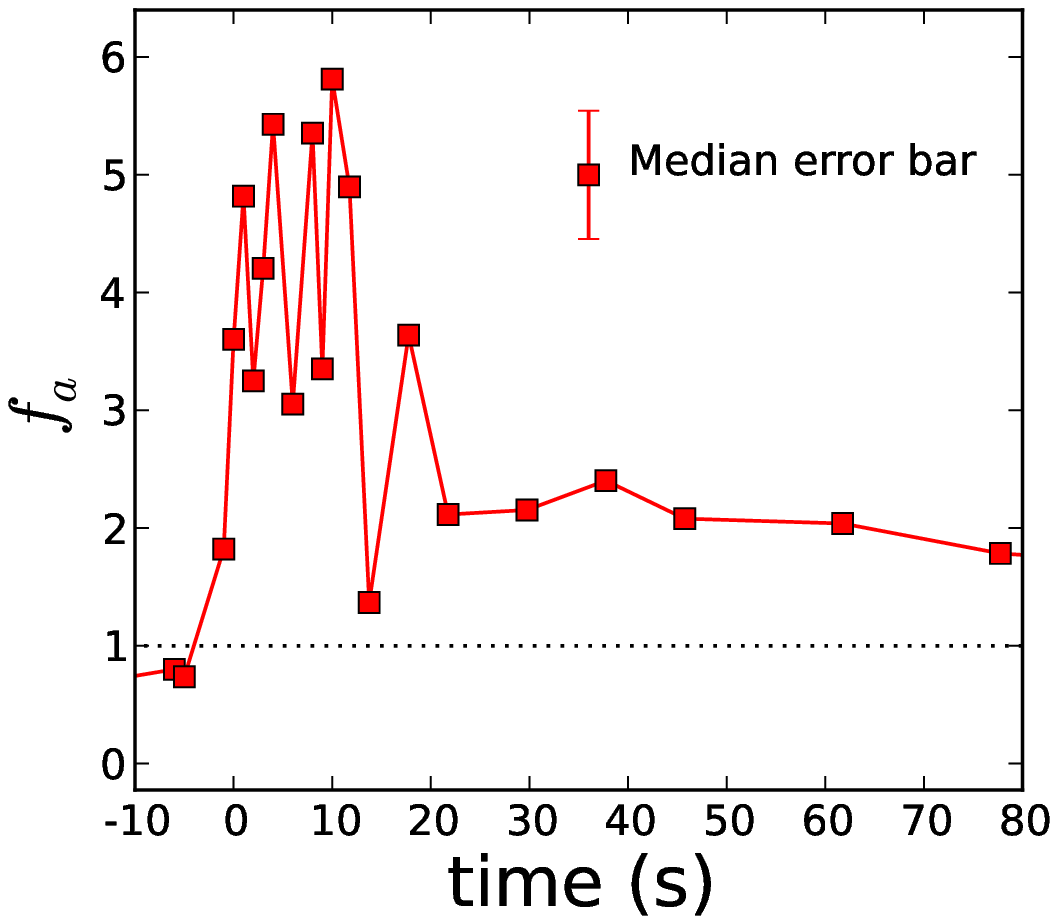}\includegraphics[width=60mm]{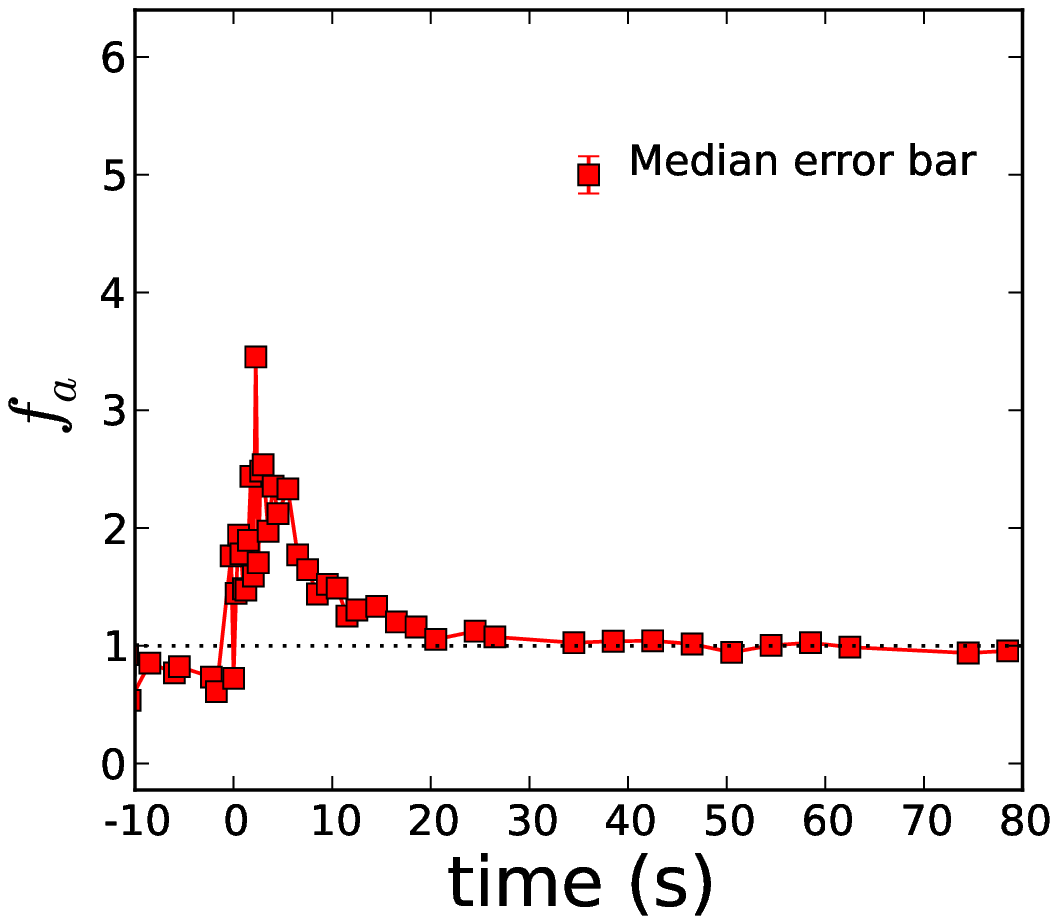}\\
\includegraphics[width=60mm]{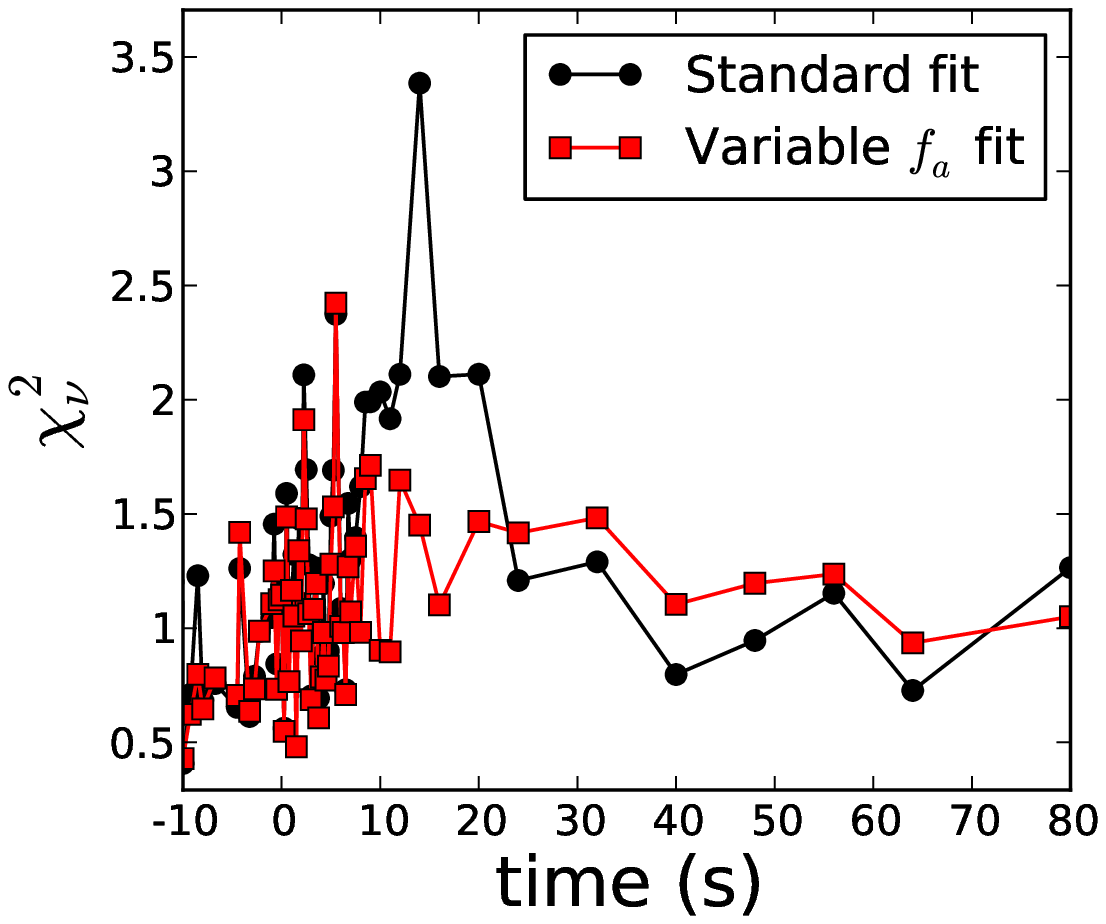}\includegraphics[width=60mm]{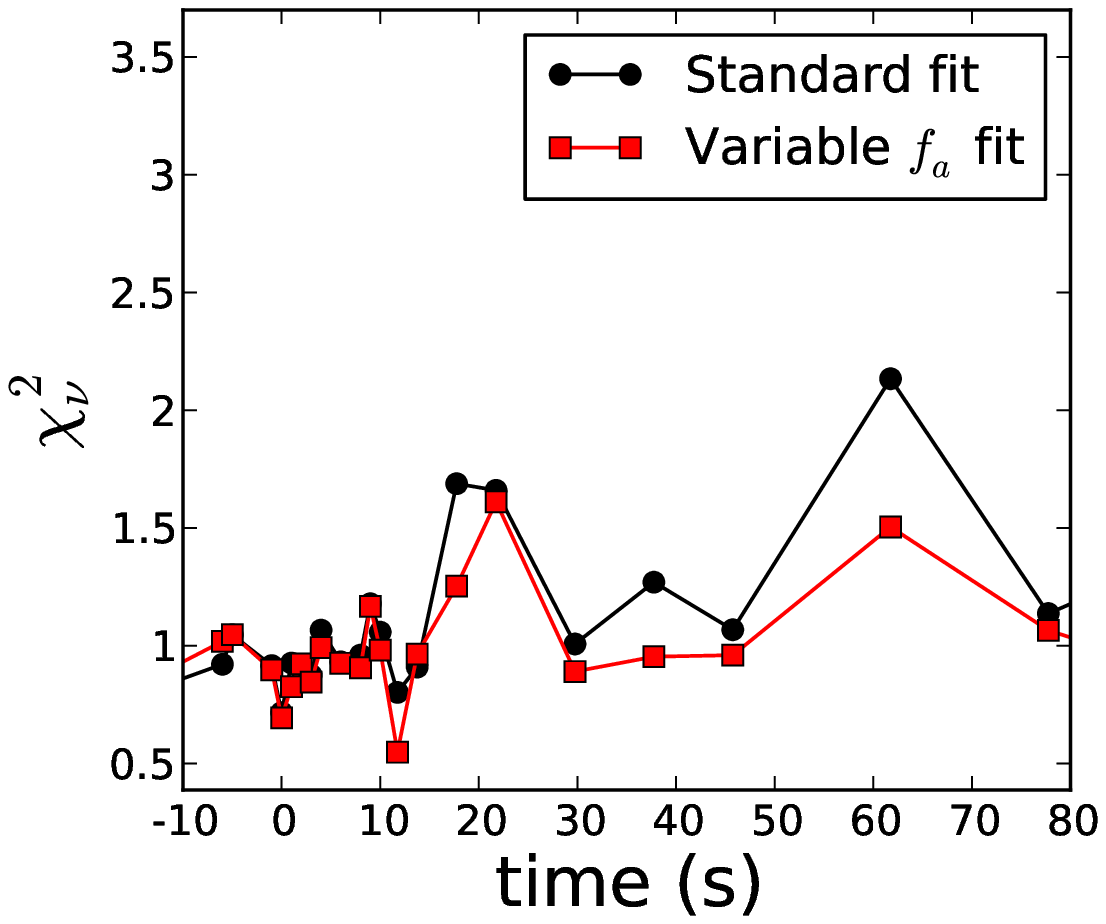}\includegraphics[width=60mm]{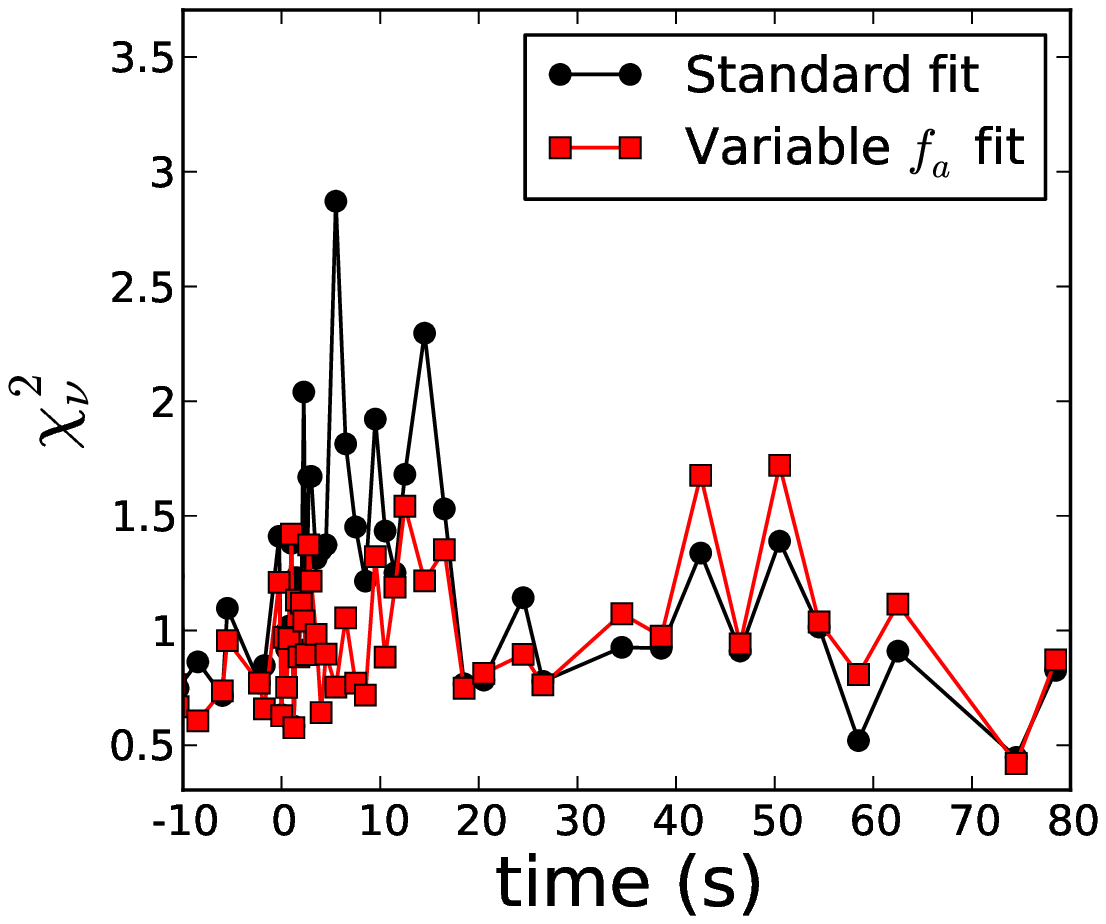}
\caption{ Comparison of fitting a variable persistent emission factor, $f_a$ , to the standard approach fits for non-PRE bursts from three different neutron stars. We have selected a typical hydrogen accretor (4U~1608$-$52), a dipping source (EXO~0748$-$676), and a He-accretor and likely ultracompact binary (4U~1728$-34$). The variable fit approach yields consistently lower unabsorbed burst component fluxes (top panels). The contribution to the flux from the variable persistent emission increases to several times the pre-burst level (middle panels) during the rise of the burst. Allowing the persistent emission to vary improves the $\chi^2_\nu$ (lower panels).}
\label{fig:layered}
\end{figure*}

Profiles of fit parameters, using both methods, for three example bursts are shown in Figure \ref{fig:layered}. We have selected bursts from three different sources to illustrate the results. These plots clearly show that $f_a$ is enhanced to several times the pre-burst level, as found for radius expansion bursts by \citetalias{WorpelEtAl2013}. This result is typical for all the bursts in our analysis. The $f_a$ for spectra preceding the burst appear to be slightly lower than unity, but this is an artefact of fitting these spectra with a model that includes a burst component which in reality is absent before the onset of nuclear burning. This causes some of the persistent flux to be misidentified as burst emission. Fitting the pre-burst spectra with just a normalization-variable persistent model gives results consistent with unity.

\begin{figure*}
\includegraphics[width=80mm]{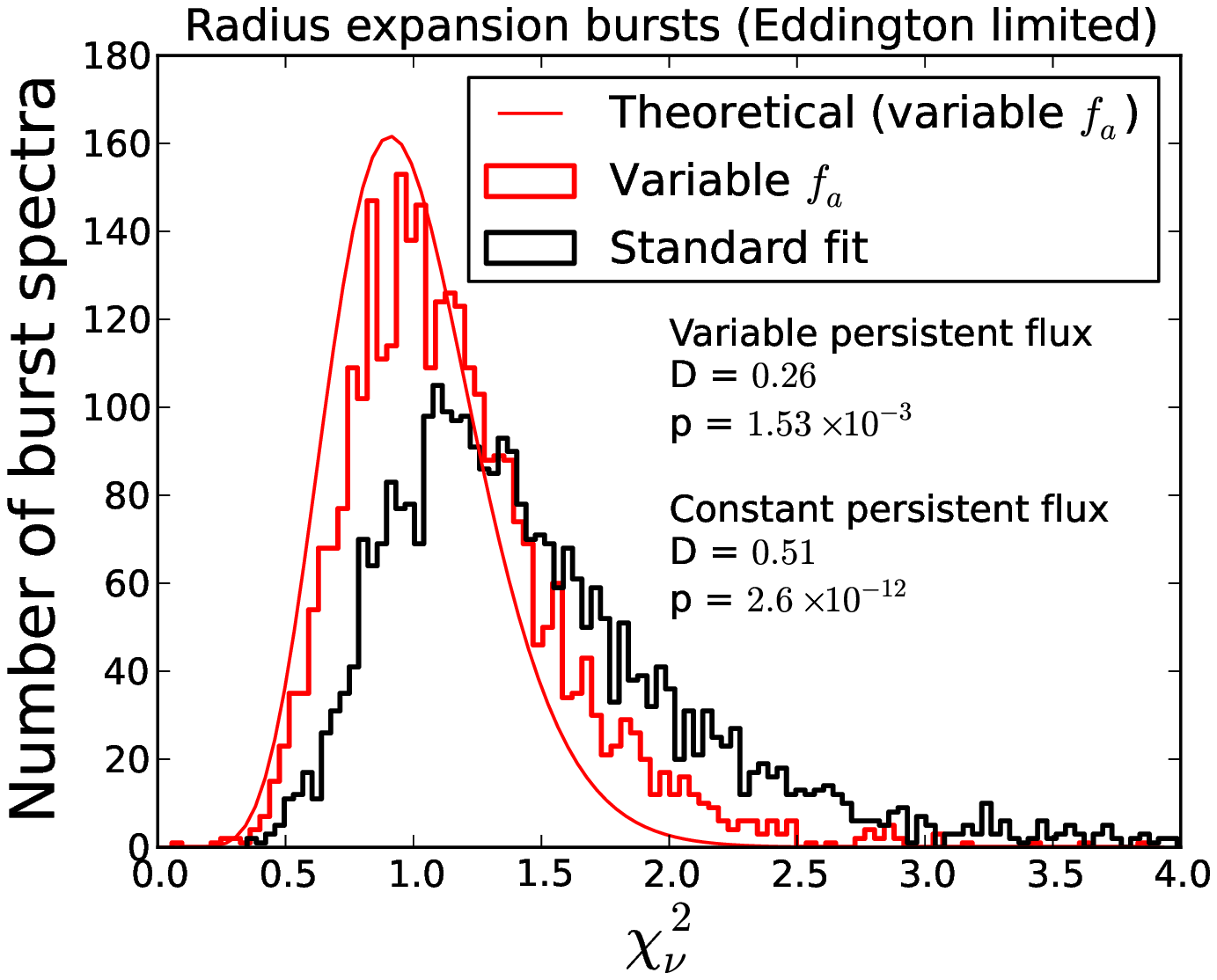}
\includegraphics[width=80mm]{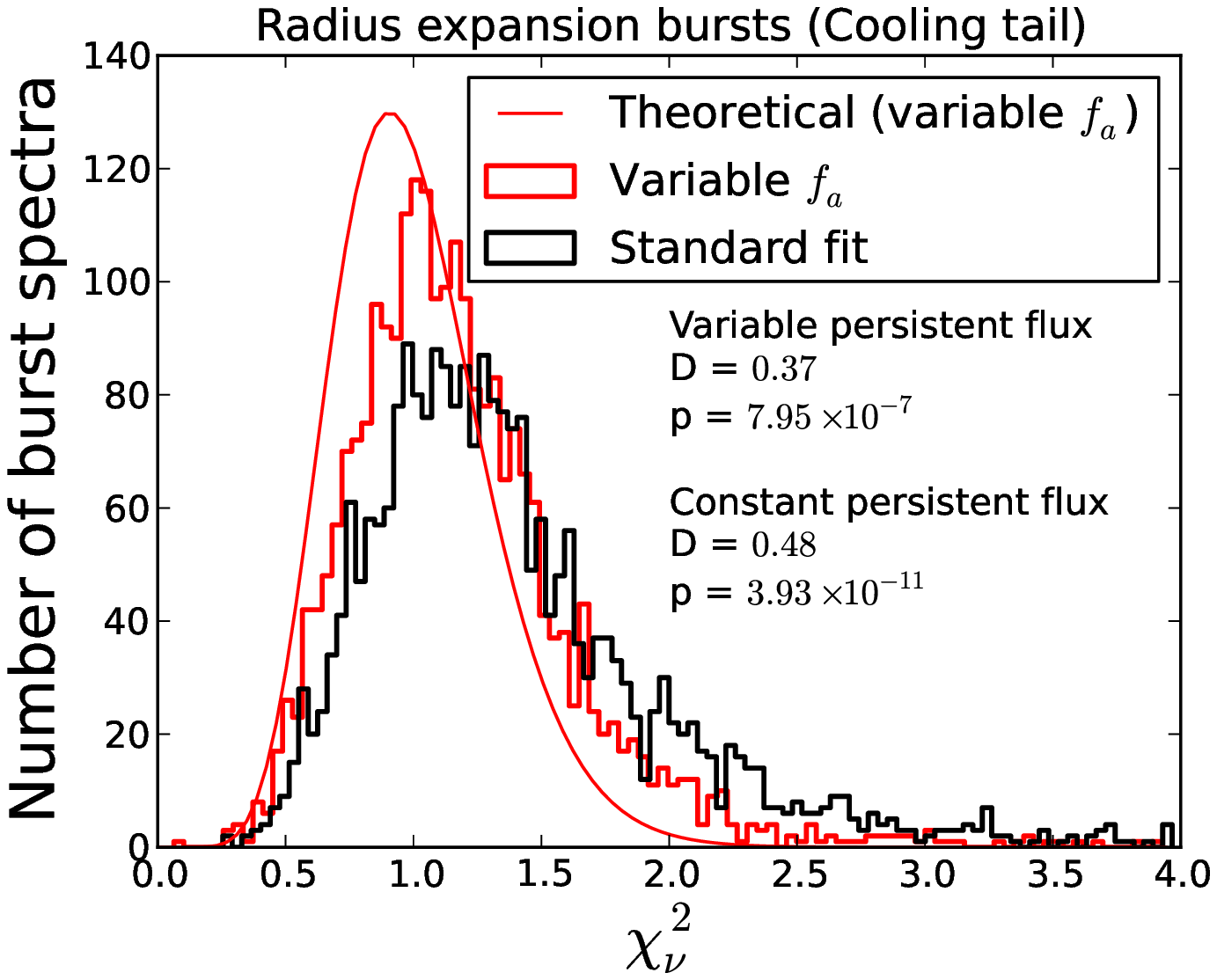}
\includegraphics[width=80mm]{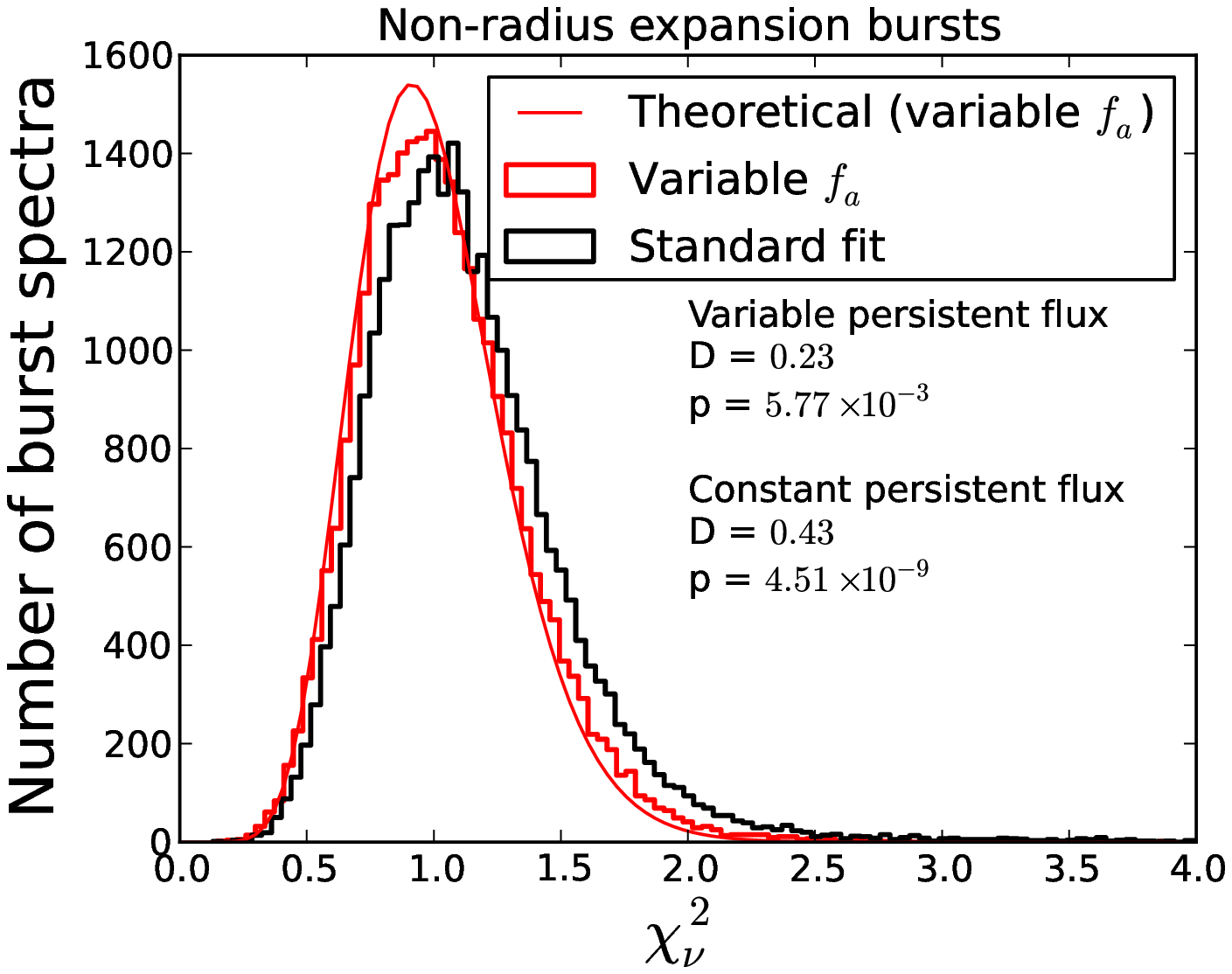}
\caption{ Histograms of the $\chi^2_\nu$ for the variable persistent normalization and standard approach spectral fits (red and black histograms respectively), and the theoretical $\chi^2_\nu$ distributions for a model that adequately describes the data (red and black curves respectively). The first panel shows spectra from Eddington-limited PRE bursts, the second panel shows cooling tail spectra from PRE bursts, and the third panel shows spectra from non-PRE bursts. It is obvious that the variable persistent normalization approach gives better fits overall than the standard approach. It can also be seen that a blackbody burst spectrum is far from acceptable in PRE bursts, even in the cooling tail when the atmosphere has returned to the surface of the star. The variable persistent normalization approach corrects for much of this effect. Non-PRE spectra are better described by a blackbody and constant persistent model, but even these are significantly improved with the variable persistent normalization approach. Results of Kolmogorov-Smirnov tests are also shown; we list the K-S statistic $D$ and null hypothesis probability $p$ for the variable persistent normalization and standard fits respectively.}
\label{fig:spec_chi2_histos}
\end{figure*}

In Figure \ref{fig:spec_chi2_histos} we show the distributions of $\chi^2_\nu$ for the variable persistent normalization fits and the standard approach fits, compared with a theoretical distribution of $\chi^2_\nu$ for a model that adequately describes the data. Fits from both PRE bursts and non-PRE bursts are shown. We performed Kolmogorov-Smirnov tests on the measured distributions of $\chi^2_\nu$ against the theoretical. $D$ and $p$ values are listed on Figure \ref{fig:spec_chi2_histos}, with the variable fits listed first. These results confirm what is visually evident in the Figure: the variable persistent normalization fits significantly improve the quality of the spectral fits for both radius expansion and non radius-expansion bursts. The spectral fits for PRE bursts are generally poorer than non-PRE bursts for both fitting methods and, although the variable persistent normalization method improves the distribution of $\chi^2_\nu$, the deviation from a model that is statistically consistent with the data is still present. It is clear that radius expansion introduces a significant spectral effect on top of the variations induced by enhanced persistent emission. The second and third panels show that cooling tail spectra of PRE bursts are more poorly fit than spectra from non-PRE bursts. This may imply that PRE affects the disc and/or photosphere in a manner that persists for some time after the photosphere has returned to the surface of the star.

We assessed the relative improvement of the variable persistent flux treatment over the standard analysis by estimating the Bayes factor, as follows. We interpret the probabilities calculated from the K-S statistics as the global likelihood for each model, i.e. $p(D|M_i,I)$ where $D$ is the set of $\chi^2$ values corresponding to the model fits, $M_i$ represents each fitting approach (model), and $I$ the priors. Even if this interpretation is not strictly correct, we expect that the K-S probability is proportional to the actual likelihood, so that the ratio of probabilities is a good estimate of the ratio of likelihoods. Following \cite{Gregory2005}, the other contribution to the Bayes factor calculation is the penalty imposed by the additional degree of freedom introduced with the variable persistent flux factor $f_a$. Since the standard approach is a special case of the variable persistent flux approach (with $f_a$ fixed implicitly at 1), we compare the effective uncertainty for the former approach (arising from the Gaussian statistics on the pre-burst persistent spectra) with the prior for the latter approach, which is flat between $f_a=-100,100$. The median total counts in all the pre-burst spectra is approximately 9,000~counts, and so we adopt $10^4$~counts, giving a characteristic width of the prior in the standard approach of $\delta f_a/f_a\equiv\sqrt{10^4}/10^4=0.01$. The overall estimate of the Bayes factor is then
\begin{equation}
B\approx \frac{p(D|M_1,I)}{p(D|M_2,I)}\frac{\delta f_a}{\Delta f_a}
 =\frac{5.77\times10^{-3}}{4.51\times10^{-9}}\times\frac{0.01}{200} = 64
 \end{equation}
Thus, even taking into account the additional degree of freedom introduced by allowing the normalisation of the persistent flux to vary, we conclude that the variable persistent flux improves on the standard approach with a Bayes factor of 64.

\begin{figure*}
 \includegraphics[width=60mm]{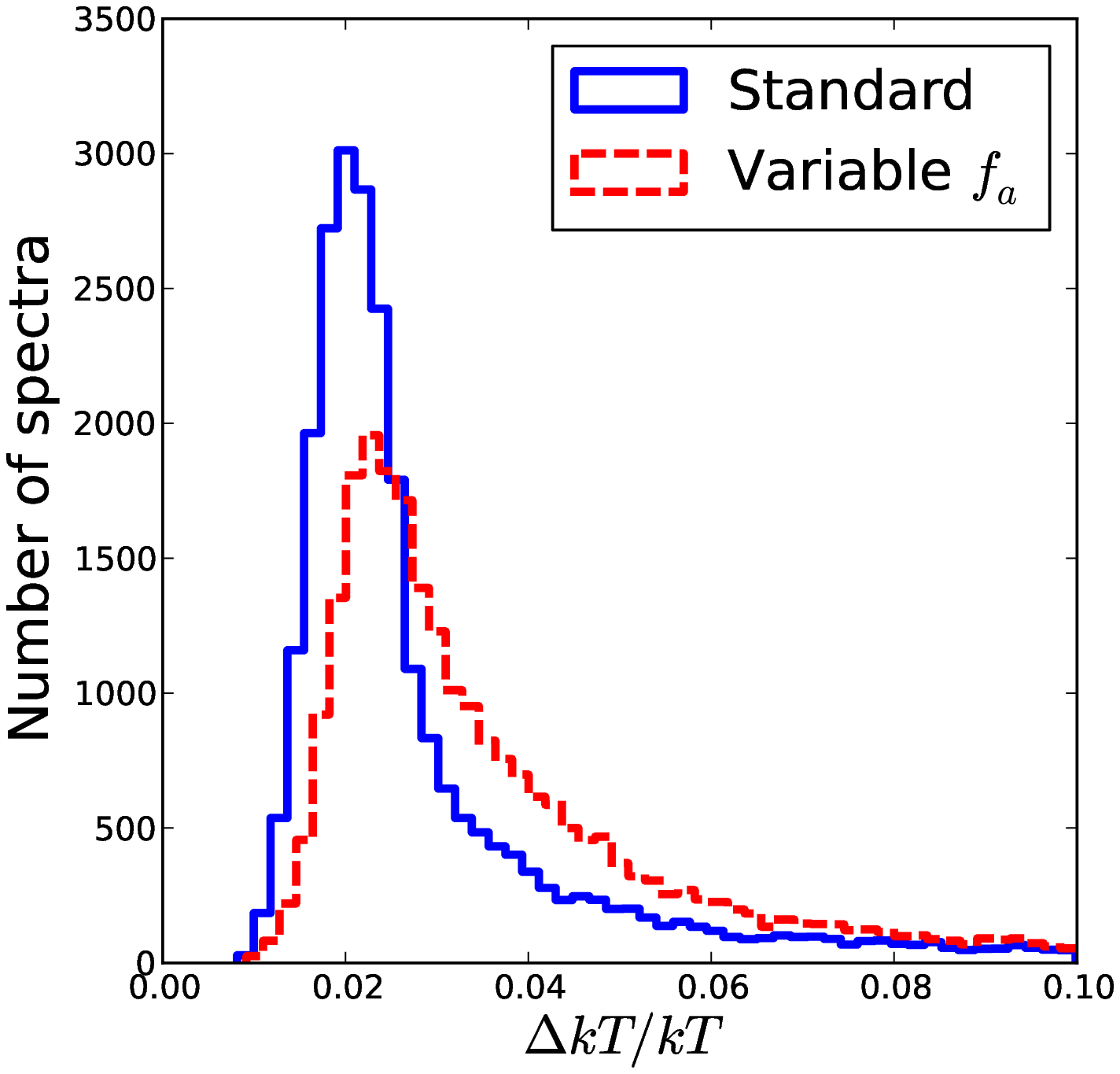}
 \includegraphics[width=60mm]{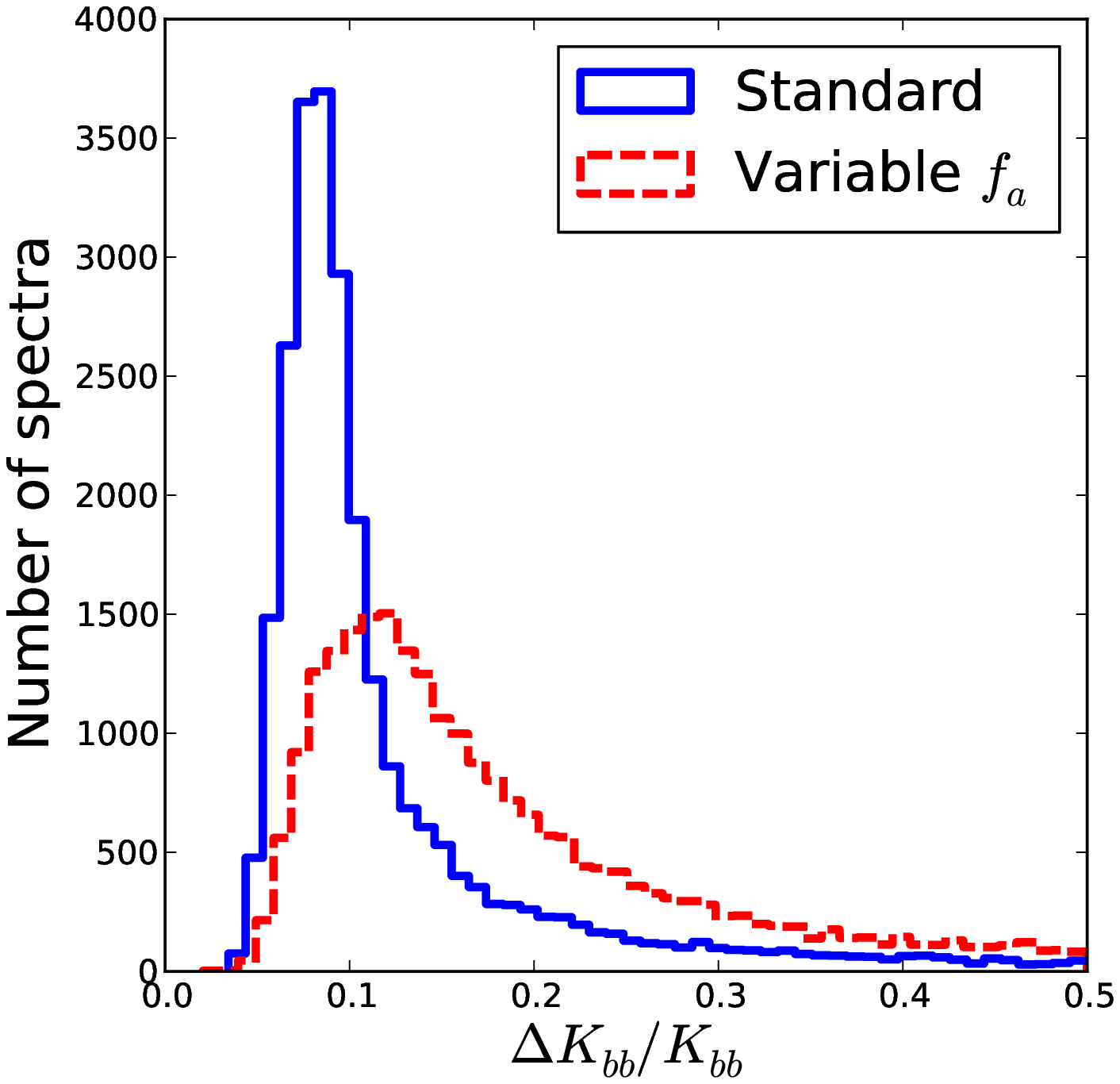}
 \includegraphics[width=60mm]{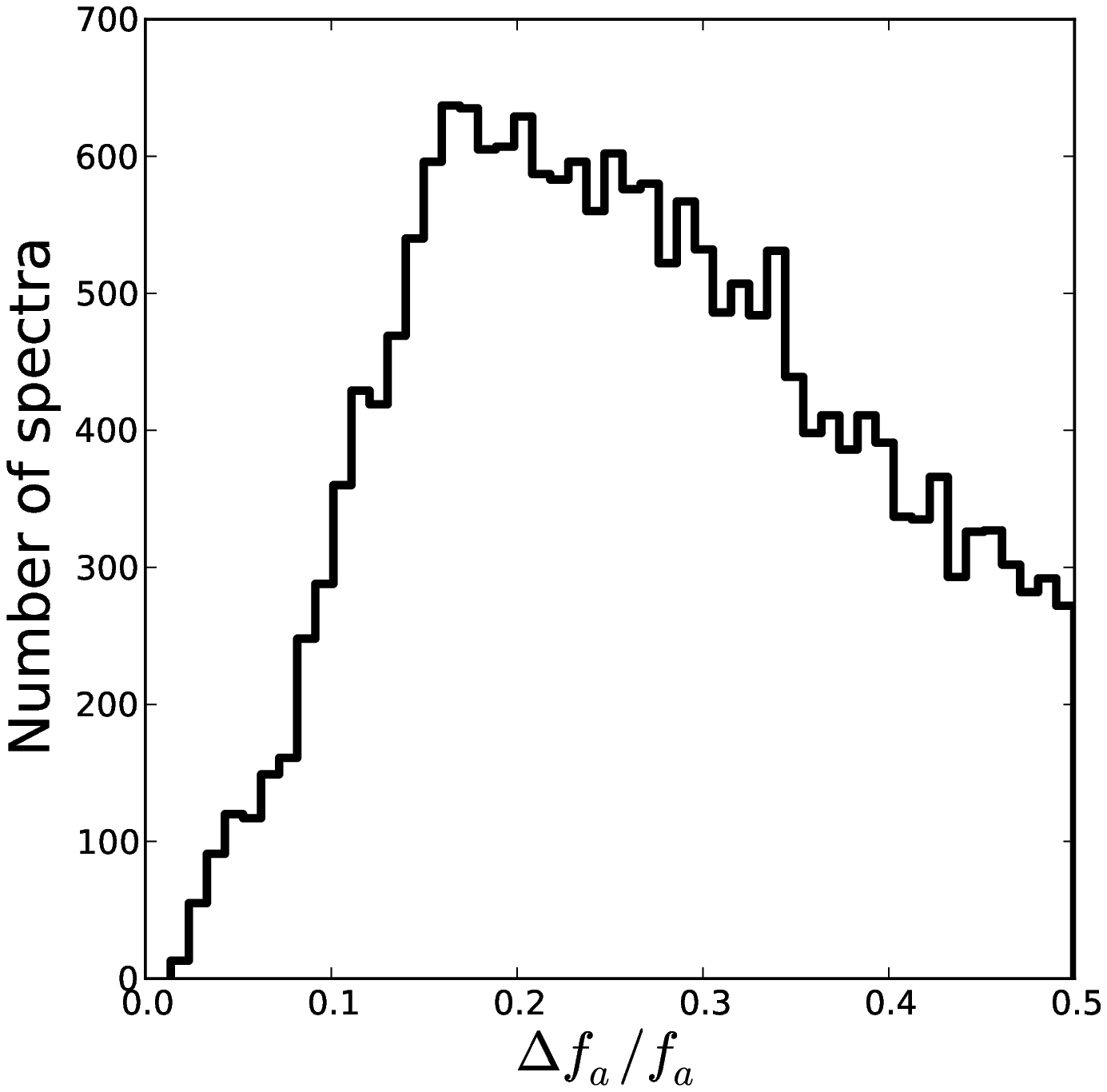}
 \caption{Histograms of the relative uncertainties in the spectral parameters using both approaches. The addition of the $f_a$ parameter causes the temperature and normalization of the blackbody component to be less tightly constrained than in the standard approach. The persistent flux factor $f_a$ itself is often very poorly constrained.}
 \label{fig:parameter_uncertainties}
\end{figure*}

The addition of a third variable parameter to the model can have an adverse effect on the uncertainties in the fit parameters. In Figure \ref{fig:parameter_uncertainties} we compare the uncertainties in fit parameters between the two approaches. It is clear that introducing $f_a$ has a slight adverse effect on the uncertainty in the blackbody temperature. The uncertainties in blackbody temperature are, on average, 25\% larger for the variable persistent flux method. The uncertainty in the blackbody normalization on average increases by 49\%, to about 18\% of the measured value. The uncertainty in $f_a$ is on average 27\% of the measured value. The poor constraints on $f_a$ and blackbody normalization are likely due to the spectral similarity of the two components. Recently \cite{BarriereEtAl2014} showed a severe example of this problem in using the variable persistent flux method in a \emph{NuSTAR} observation of GRS~1741.9$-$2853, in which the $f_a$ factor was not constrained at all. As they point out, the problem occurs when the pre-burst persistent emission is faint, and when it closely resembles the burst component in spectral shape.

\citetalias{WorpelEtAl2013} found, for PRE bursts, only a weak correlation between $f_a$ and burst flux. They found that, although most of the high $f_a$ values occurred during radius expansion, there was nonetheless great variability in $f_a$ when the burst flux was relatively constant. We have repeated this analysis for non-PRE bursts. We have taken all spectra for each burst between the beginning of the burst and the point at which burst flux declines below 10\% of its peak value, and performed Kendall $\tau$ rank correlations on burst flux against $f_a$ for each burst. We found positive correlations in 1000 out of 1419 non-PRE bursts, of which 919 are at greater than 3$\sigma$ significance. Performing the same test on the cooling tails of PRE bursts, we found positive correlations in 299 out of 316. Of these, 286 were at greater than 3$\sigma$ significance. These numbers demonstrate that there is a more direct relationship between burst flux and $f_a$ for non-PRE bursts than for PRE bursts.

\citetalias{WorpelEtAl2013} found that, in PRE bursts, the variations in measured $f_a$ cannot be attributed to counting statistics. We repeated this test for all non-PRE bursts in our catalog. For all spectra obtained during non-PRE bursts we generated 100 simulated spectra using the standard approach spectral parameters, and incorporating counting statistics typical of the detector. We then fit these simulated spectra with the variable persistent normalization method, obtaining a standard deviation of $f_a$ measurements arising from counting statistics in the absence of any variation of persistent flux. We found real $f_a$ measurements greater than unity to $5\sigma$ significance in 214 of 62783. In the absence of a real effect we would expect at most one such case arising by chance. This experiment shows that enhanced persistent emission is present in non-PRE bursts as well.

Since PRE bursts are by nature brighter than non-PRE bursts, it is possible that the difference in $\chi^2_\nu$ between the two classes is simply due to superior signal-to-noise in the brighter bursts. To investigate this we took a subset of spectra from both PRE and non-PRE bursts, with bolometric fluxes between $4.0\times10^{-8}$ and $5.0\times10^{-8}$~erg s$^{-1}$ cm$^{-2}$ and blackbody temperatures between 2.0 and 2.5~keV, these quantities being measured with the standard approach. There were 331 spectra from PRE bursts and 655 spectra from non-PRE bursts in this segment of the parameter space. These had mean $\chi^2_\nu$ of $ 1.17$ and $ 1.35$ for non-PRE and PRE bursts respectively, and a K-S test indicated a probability of $ 1.79\times 10^{-6}$ that these $\chi^2_\nu$ are drawn from the same distribution. This shows that the poorer spectral fits generally observed in PRE bursts are not due to better signal-to-noise in the bright bursts.
\section{Is the persistent emission spectral shape constant?}
\label{sec:pers_shape}
Our approach of treating the persistent emission as a contribution of fixed shape depends upon the assumption that its shape does not change, at least not on timescales of the order of the burst duration, or to a degree sufficient to affect our spectral fits. \citetalias{WorpelEtAl2013} did not investigate this matter directly, but their finding that $f_a$ generally returns to its pre-burst level, with good $\chi^2_\nu$, suggests that this assumption is not unreasonable. Similarly, \cite{KeekEtAl2014} and \cite{PeilleEtAl2014} also report that the persistent emission returns to its original level soon after the burst. \cite{ThompsonEtAl2005} and \cite{BagnoliEtAl2013} have found that the persistent emissions of 4U~1826$-$24 and the Rapid Burster respectively are remarkably stable over kilosecond timescales between bursts. \cite{LinaresEtAl2014} found variability over $\sim$10~minute periods in the \emph{intensity} of the persistent emission of IGR~J18245$-$2452, with no change in the \emph{shape} of the persistent emission. Dipping systems, such as EXO~0748$-$676 and X~1658$-$298, show somewhat more variability \citep{SidoliEtAl2005, OosterbroekEtAl2001} on $\sim$1~m timescales, and the spectrum becomes harder during dips, as indicated by hardness ratios. However, the dips are due to obscuration in these high-inclination systems. Outside of dipping intervals, the hardness ratios are effectively constant. It is therefore likely that no change in the shape or intensity of the persistent emission is to be expected in the absence of burst luminosity on minute to hour timescales.

\begin{figure}
\includegraphics[width=80mm]{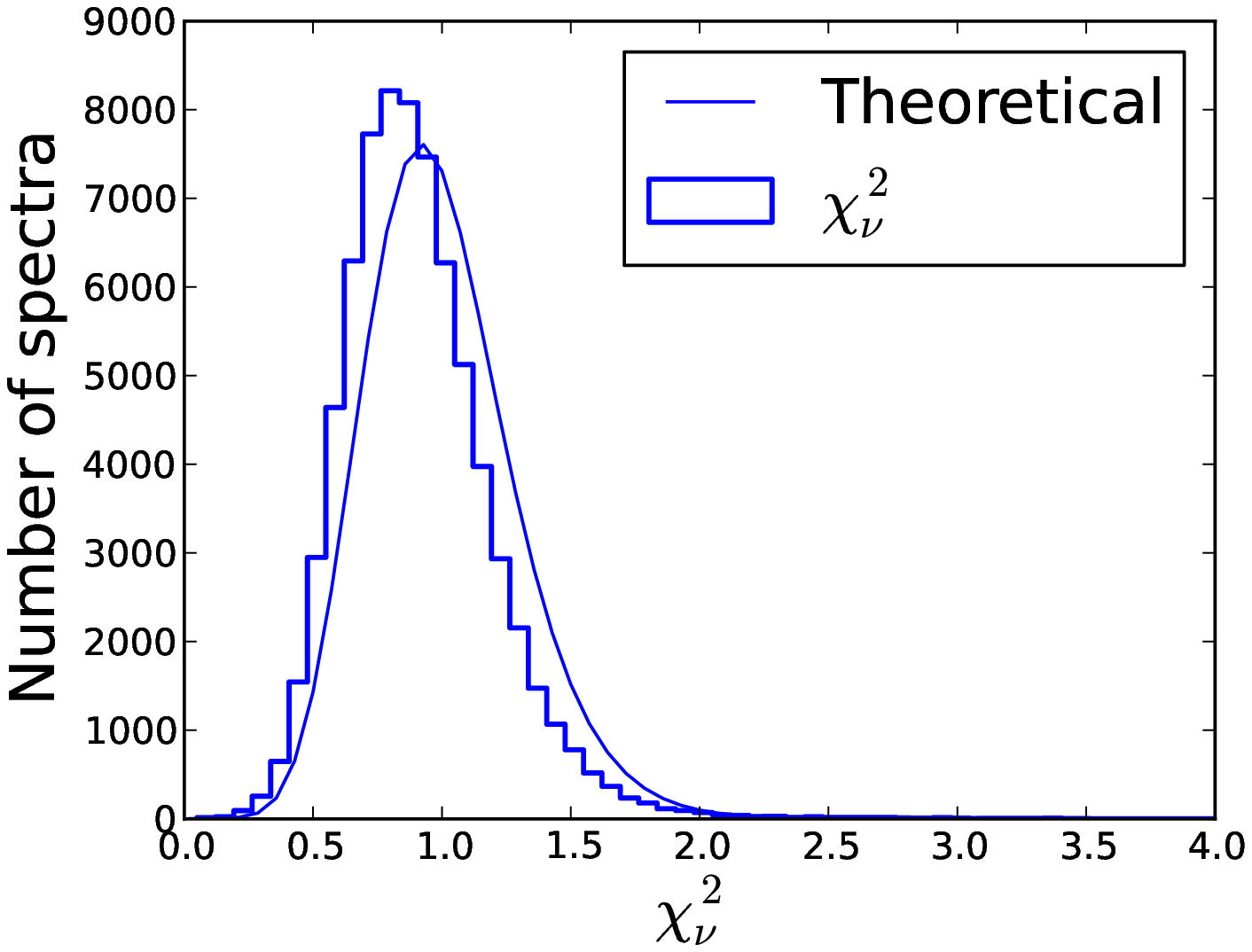}
\includegraphics[width=80mm]{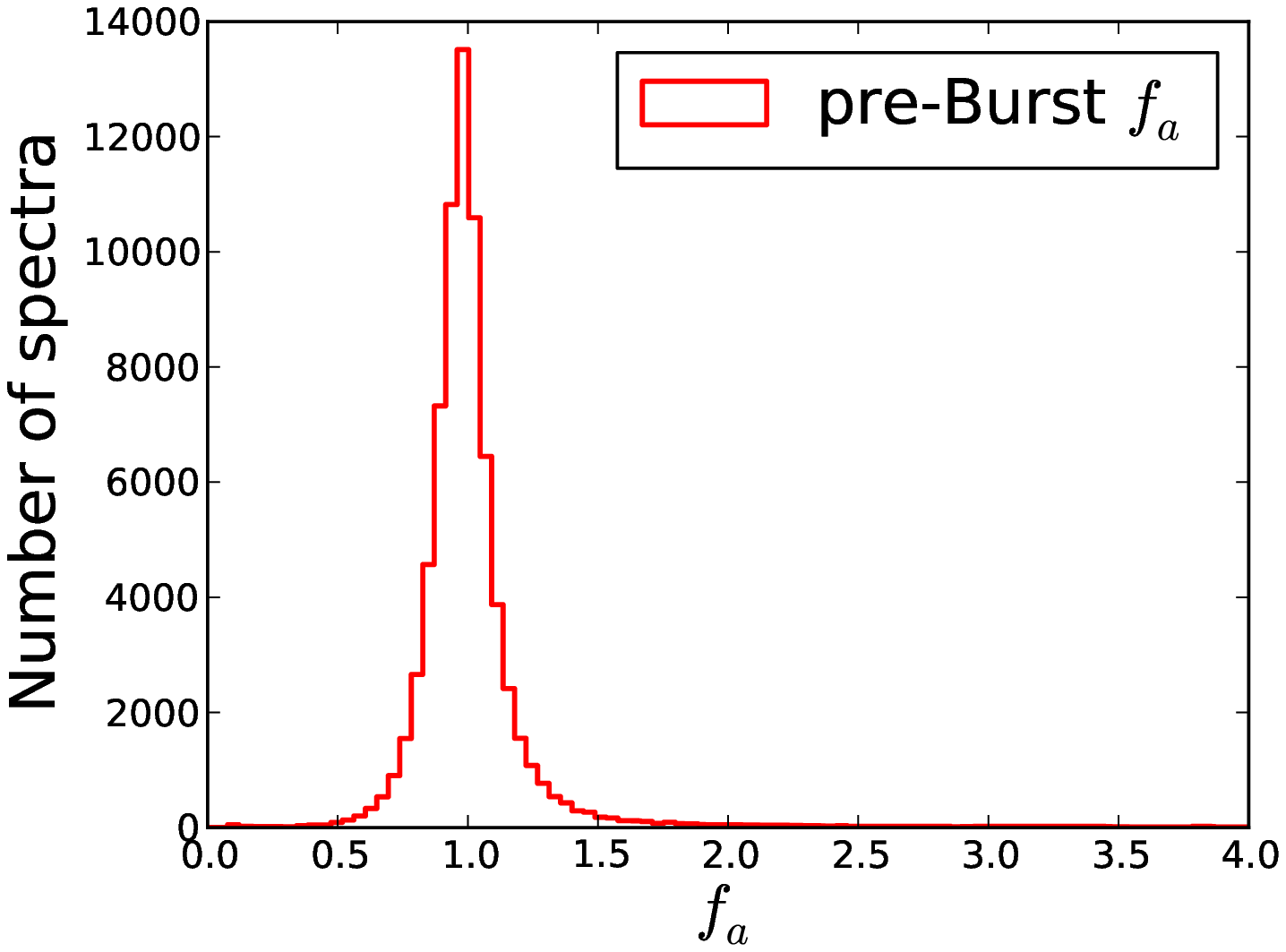}
\caption{Measured distribution of $f_a$ (red histogram), and measured and theoretical $\chi^2_\nu$ (thick and thin blue curves) for pre-burst emission in the absence of any nuclear burning. The $f_a$ values are very strongly peaked around 1, and the measured $\chi^2_\nu$ are low, indicating that the pre-burst emission does not change intensity or spectral shape to any detectable degree.}
\label{fig:pers_variability}
\end{figure}

On shorter timescales, the pre-burst light curve is remarkably flat preceding bursts for 4U~1636$-$536 \citep{NathEtAl2002}, the Rapid Burster \citep{BagnoliEtAl2013}, and 4U~1608$-$52 \citep{PeilleEtAl2014}, and is also flat over a large range of accretion rates in IGR~J17480$-$2446 \citep{MottaEtAl2011}. These works suggest that short-term intensity variations are not prevalent in the persistent emission preceding bursts. To gain further confidence that the persistent emission shape and intensity are not significantly variable on shorter timescales than minutes or hours, in the absence of a burst, we took every 0.25s pre-burst spectrum from every burst and fit it with just $f_a$ times the persistent model for that burst, i.e., no thermonuclear burst emission. The distributions of $f_a$ and $\chi^2_\nu$ are shown in Figure \ref{fig:pers_variability}. This data typically covers a few tens of seconds prior to every burst. It is plain to see that $f_a$ is distributed tightly around unity, indicating that the persistent emission generally does not vary greatly in intensity on timescales of 1~s or less. The $\chi^2_\nu$ distribution also implies that the shape of the persistent emission is not changing appreciably, although of course the short exposure times and associated low photon counts make detecting such changes difficult.

\begin{figure}
\includegraphics[width=80mm]{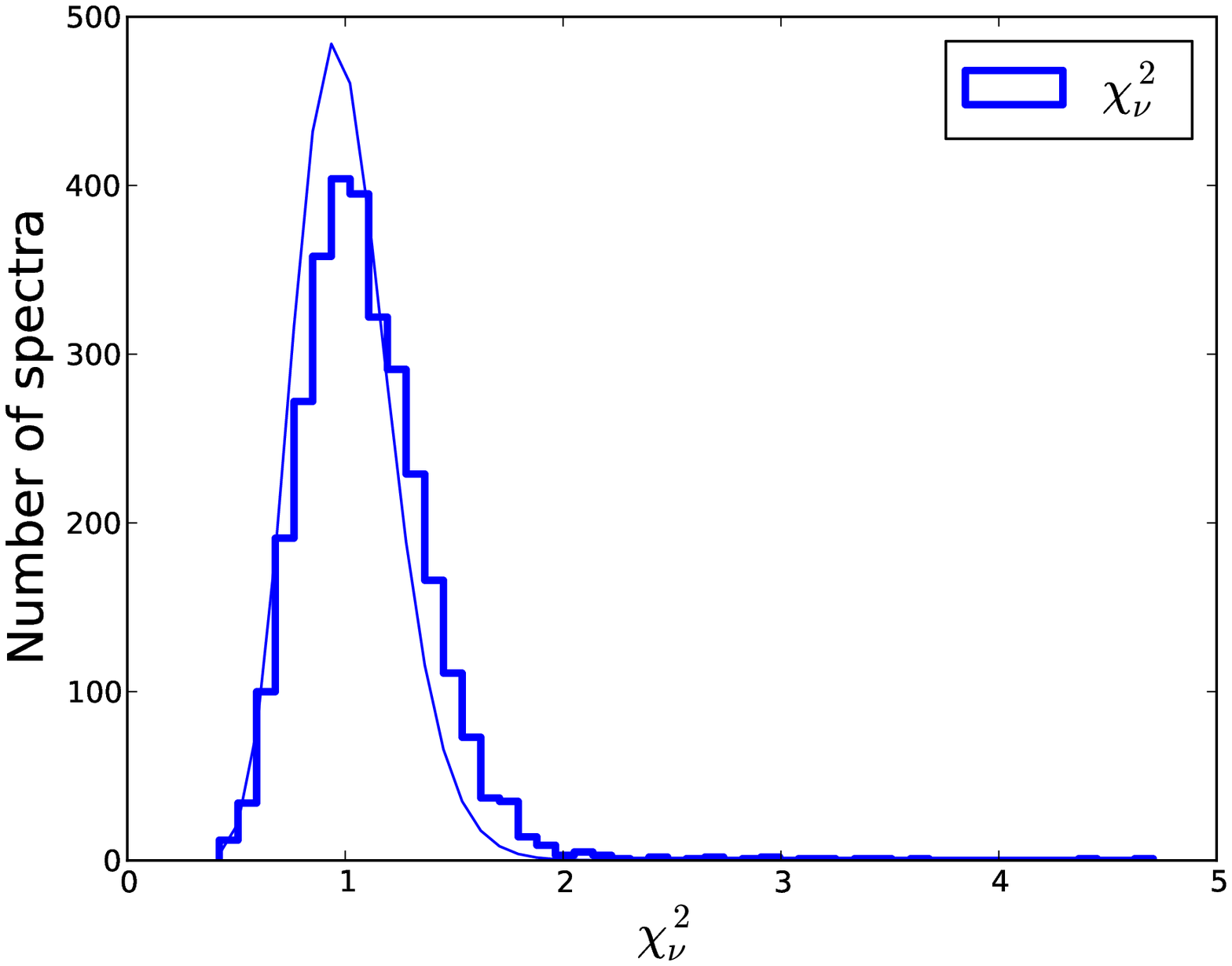}
\includegraphics[width=80mm]{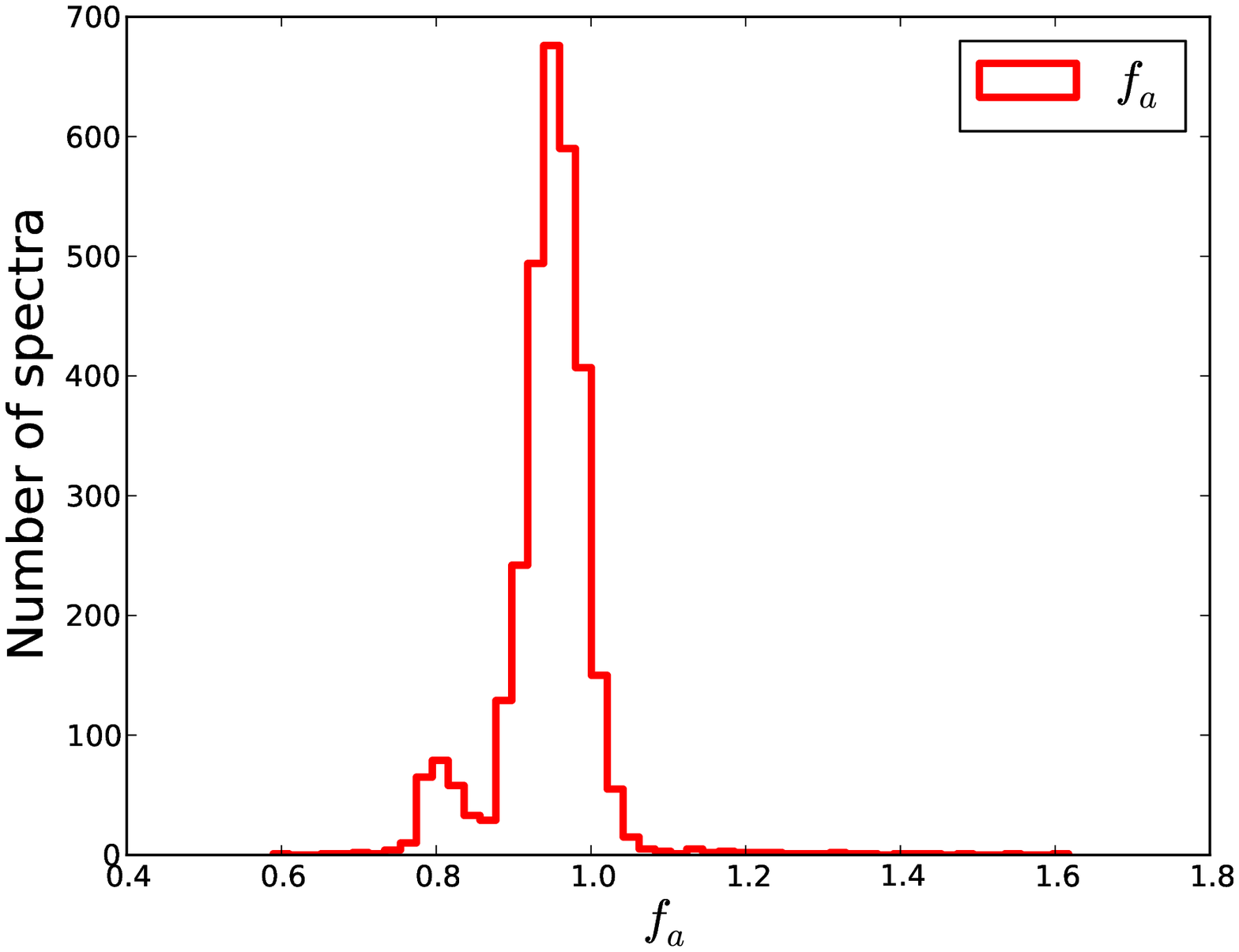}
\caption{Measured distribution of $f_a$ (red histogram), and measured and theoretical $\chi^2_\nu$ (thick and thin blue curves) for Standard-2 data covering the observations of eighteen selected bursts. The $f_a$ values are tightly clustered around unity, and the $\chi^2_\nu$ are marginally higher than the theoretical distribution assuming a well-fitting model. This suggests that the persistent emission does not change shape or intensity to any detectable degree on timescales of an hour before and after a type I burst.}
\label{fig:std2_distros}
\end{figure}

To test for more subtle spectral shape changes, we increased the exposure time to obtain better signal-to-noise. For the three bursts studied in Figure \ref{fig:layered}, and the fifteen bursts investigated by \cite{PeilleEtAl2014} (see their Table 1), we inspected the \RXTE\ Standard-2 data covering the entire observation. These are spectra binned into 16s intervals.  We fit each of these spectra with the persistent model for that burst, with all parameters frozen, and $f_a$ normalization allowed to vary. We considered only spectra recorded within an hour before or after the burst but excluding those \emph{during} the burst, to avoid contamination by the burst emission. That is, we excluded spectra recorded between the beginning of the burst (burst flux has reached 1/4 of the peak burst flux) and the end of the PCA data of the \citetalias{GallowayEtAl2008} catalog. This data typically covers a few minutes to an hour before and after the burst. We also excluded an eclipse of EXO~0748$-$676 beginning at 2004-09-25 07:20:29, 15 minutes before the burst, when no flux was observed from the neutron star (see also \citealt{HomanEtAl2003}), and spectra occurring after a data gap towards the end of an observation of 4U~1636$-$536 that began at 2001-09-05 04:56:51. We analyzed the distribution of $f_a$ and $\chi^2_\nu$. These are plotted in Figure \ref{fig:std2_distros}, showing that almost all of $f_a$ values measured this way are clustered tightly around $f_a=0.95-1.0$. There is a smaller peak at around $f_a\approx 0.85$, dominated by two bursts, but visual inspection of their $f_a$ curves show their $f_a$ values to stay approximately constant throughout the observation. 

The steadiness of both $f_a$ and $\chi^2_\nu$ over time suggests that the persistent emission does not change greatly in either shape or intensity on time scales up to an hour before and after a type I burst.

Finally, we investigated the extent to which we can detect changes in the accretion spectral shape if they really are present, assuming that a change in the intensity of the persistent emission reflects a change in the accretion rate. This is particularly necessary near the burst peak, since this is where the persistent emission spectral shape is most uncertain; we have already found that it returns to its original shape deep in the cooling tail. We selected a non-PRE burst from 4U~1636$-$536 that uses the \texttt{disko} model \citep{StellaRosner1984}, recorded 2005 Apr 4, 10:48:43. For this burst PCUs 0 and 2 on \RXTE\ were active. This model contains $\dot{M}/\dot{M}_\text{Edd}$ explicitly as a variable parameter and incorporates physically-motivated changes in the spectral shape. This model is assumed to be in equilibrium for this burst and does not take into account changes in its structure induced by a burst; there are as yet no theoretical spectra of a disk modified in this way, and we expect that the associated spectral shape will deviate from a blackbody to a similar degree, though not perhaps in the same manner.

\begin{figure*}
\includegraphics[width=50mm]{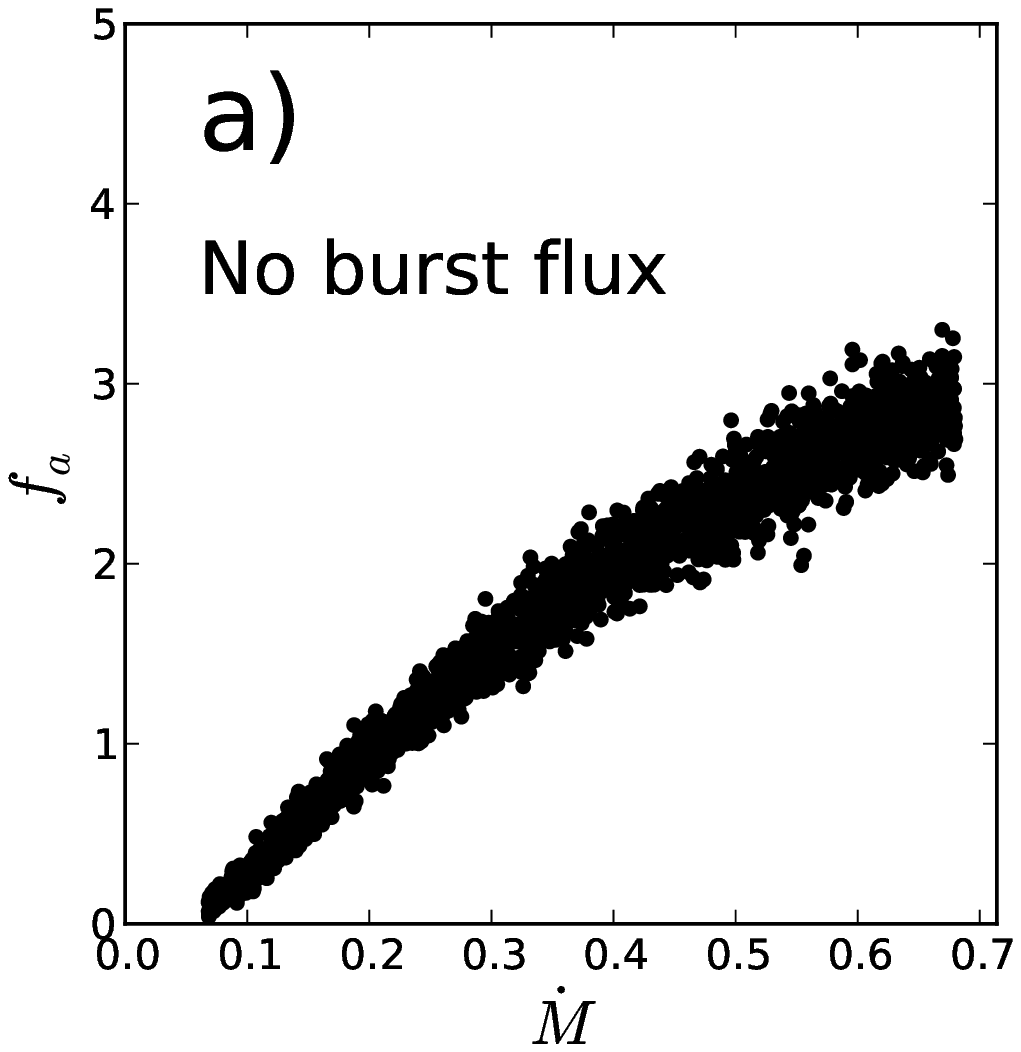}
\includegraphics[width=50mm]{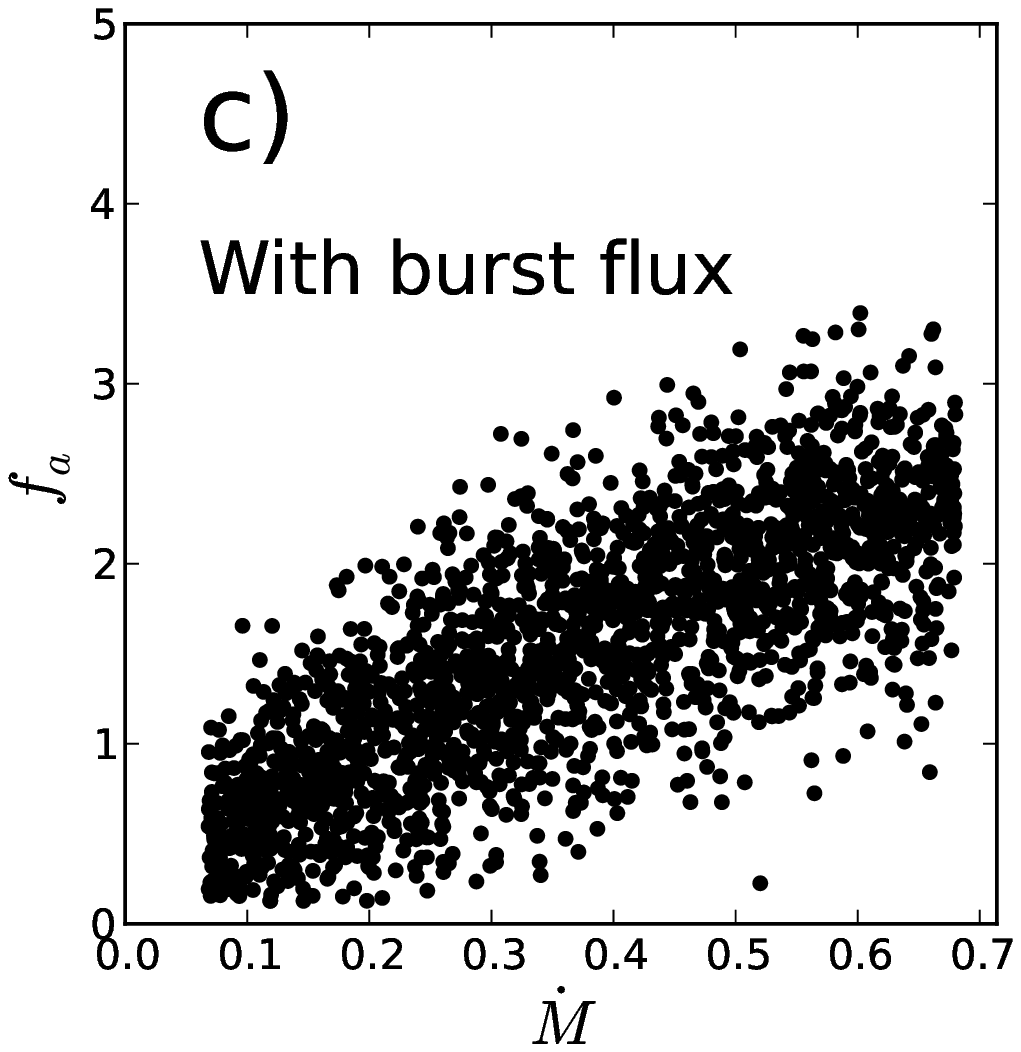}\\
\includegraphics[width=50mm]{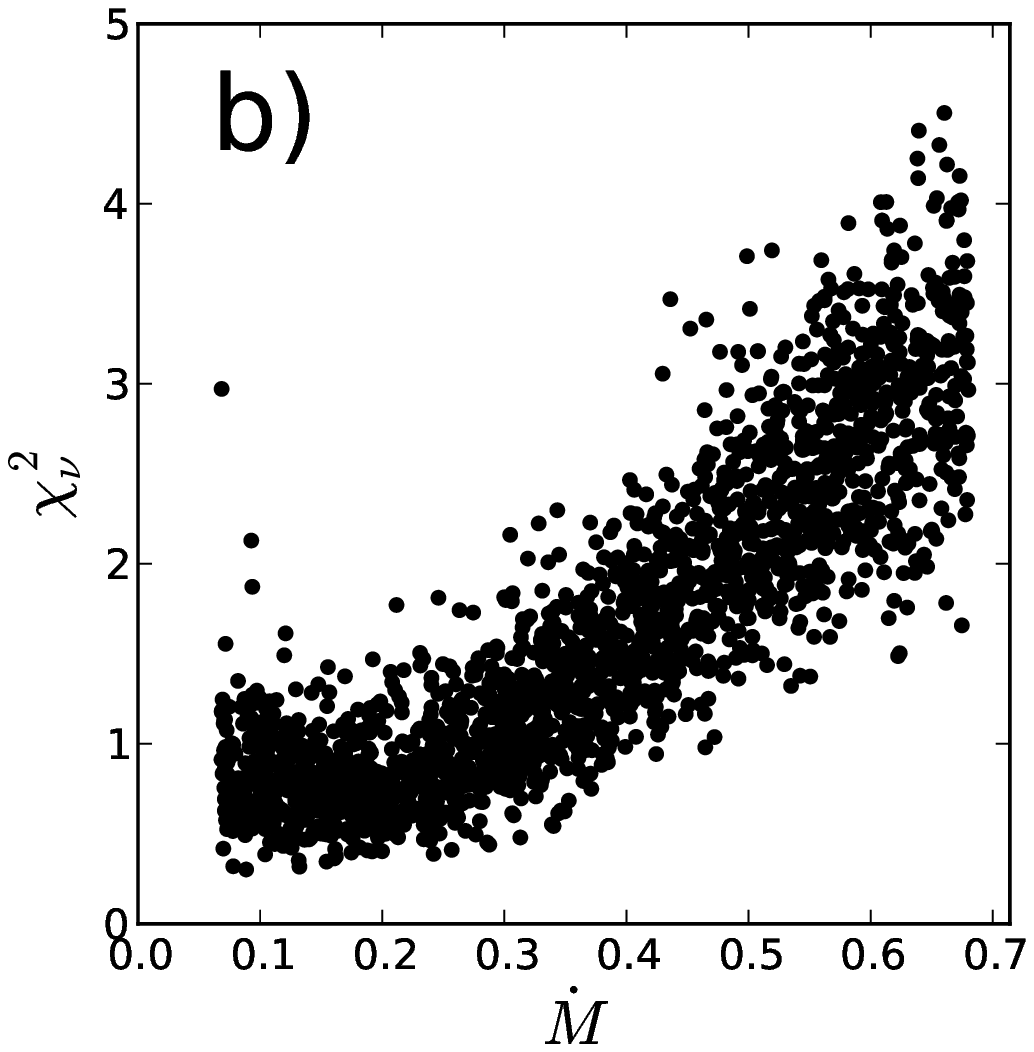}
\includegraphics[width=50mm]{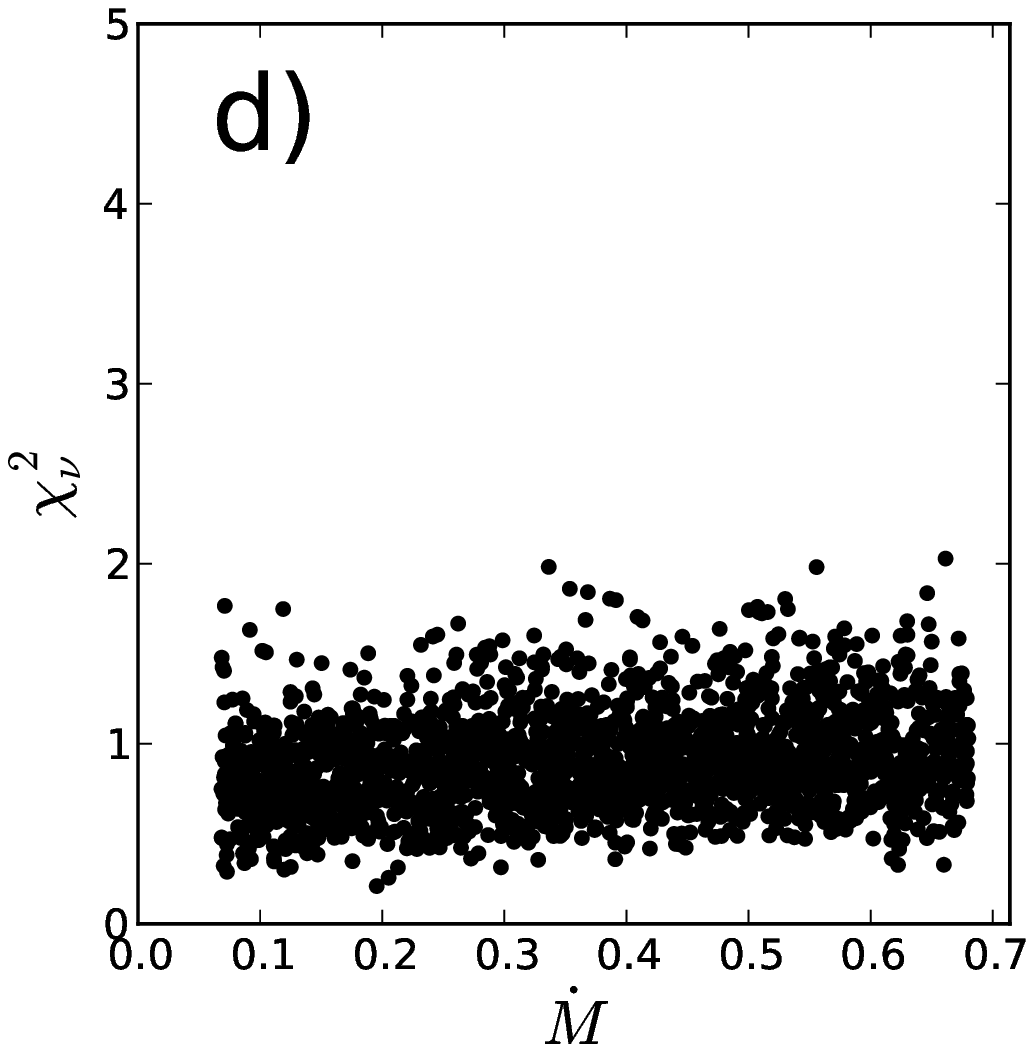}\\
\caption{ Results of fits to simulated \texttt{disko} spectra with $\dot{M}$ allowed to vary, both with (panels a. and b.) and without (panels c. and d.) a burst component. In both cases an increase in $f_a$ is detected. When there is no burst component, and only a change in the intensity of the persistent emission is being tested for, the increasing $\chi^2_\nu$ with increasing $\dot{M}$ indicates that changes in the spectral shape are detectable. In the presence of burst emission, it is no longer possible to discern the spectral shape changes in the persistent emission. We can therefore simply vary the normalization of the persistent emission without introducing systematic effects.}
\label{fig:mdotfakes}
\end{figure*}

The best fit accretion rate was 0.205$\dot{M}/\dot{M}_\text{Edd}$ and the peak $f_a$ was 3.3. The normalization of this persistent model was $2\cos i/d^2=1.648$, where $i$ is the inclination and $d$ is the distance in units of 10~kpc. At an assumed distance of 6~kpc \citep{GallowayEtAl2006}, this gives an inclination of 72$^\circ$, in agreement with \cite{PandelEtAl2008} and \cite{CasaresEtAl2006}, who also infer a high inclination (greater than $64^{\circ}$, and $36^{\circ}-74^{\circ}$ respectively) for this system. The normalization was kept frozen in the subsequent analysis. We generated 2,000 simulated spectra, folded through a PCA response appropriate to the observed burst, of the \texttt{wabs*disko} model with $\dot{M}$ taking random values between 0.067$\dot{M}_\text{Edd}$ and 0.667$\dot{M}_\text{Edd}$. We then fitted those simulated spectra with $f_a$ times the original persistent model. The measured $f_a$ and $\chi^2_\nu$ are given in Figure \ref{fig:mdotfakes}, left two panels. In the absence of burst emission we can clearly detect a change in spectral shape via the increase in $\chi^2_\nu$ with increasing $\dot{M}$ as well as an increase in the intensity of the persistent emission. We then repeated the process, this time including a blackbody with temperature 2.0~keV and normalization representing a sphere with radius 10~km at a distance of 6~kpc (i.e., the surface of 4U~1636$-$536). This we fit with the usual variable persistent flux approach (right two panels). It is evident that, although we can still detect enhanced persistent emission through an increased $f_a$, we have lost the ability to detect a change in its spectral shape. This experiment gives us confidence that our approach of simply varying the normalization of the persistent emission will introduce no detectable systematic effects.

This analysis is not sensitive to changes in the persistent spectrum outside the 2.5-20~keV energy range we consider in this paper. \cite{intZandEtAl2013} combined data from Chandra, which is sensitive at energies below 2keV, with \RXTE\ data for a radius expansion burst from SAX~J1808.4$−$3658. They obtained similar $f_a$ values as \citetalias{WorpelEtAl2013} did studying the same burst with only \RXTE\ data, suggesting that changes in the persistent spectrum below 2.5keV are negligible. At high energies, there is evidence that the hard X-ray (>30 keV) flux decreases during a burst, in apparent conflict with our result \citep{ChenEtAl2013,JiEtAl2013,JiEtAl2014,JiEtAl2014b}. However, this phenomenon is attributed to the rapid cooling of the accretion disc corona, which contributes the majority of the very hard X-rays. An increase in 2.5-20keV flux attributed to accretion rate change does not conflict with a simultaneous quenching of high energy photons through the cooling of the corona.

\section{Eddington fluxes from non-PRE bursts}
\label{sec:Edd_from_nonpre}

For neutron stars for which no radius expansion bursts have been observed, the only lower bound on their Eddington fluxes are the peak fluxes of their brightest bursts. A better knowledge of their Eddington fluxes would be of observational utility, as many of these sources show interesting properties: the prolific burster GS~1826-24 exhibits very regular bursts that are in close agreement with numerical models of X-ray bursts (e.g., \citealt{HegerEtAl2007, GallowayLampe2012}). The 11~Hz pulsar IGR~17480$-$2446 is the most slowly rotating source that shows burst oscillations \citep{CavecchiEtAl2011}, making that source an important test of burst oscillation models. Neither source has a single reported PRE burst. The Rapid Burster (MXB~1730-335) is our major source of knowledge about type II bursts, and has one reported PRE burst \citep{SalaEtAl2012}. This event, however, was detected by the \emph{Swift} X-ray instrument and the measurement was not sensitive enough to constrain the Eddington flux to better than about 50\% (see their Fig. 2).

An interesting observation is that all the results of \citetalias{WorpelEtAl2013} fall below the line $f_a \gamma=1$, where $\gamma$ is the pre-burst accretion flux as a fraction of the Eddington flux, consistent with predictions of \citet{BurgerKatz1983} and \citet{MillerLamb1996}; both works suggest that the Eddington accretion rate is a natural upper limit to $\dot{M}$. This observation raises the possibility of obtaining a lower bound on the Eddington flux for a given source, that is, $F_\text{Edd}>f_aF_\text{pers}$, where $F_\text{Pers}$ is the bolometric flux of the pre-burst persistent emission.

\begin{figure*}
\includegraphics[width=150mm]{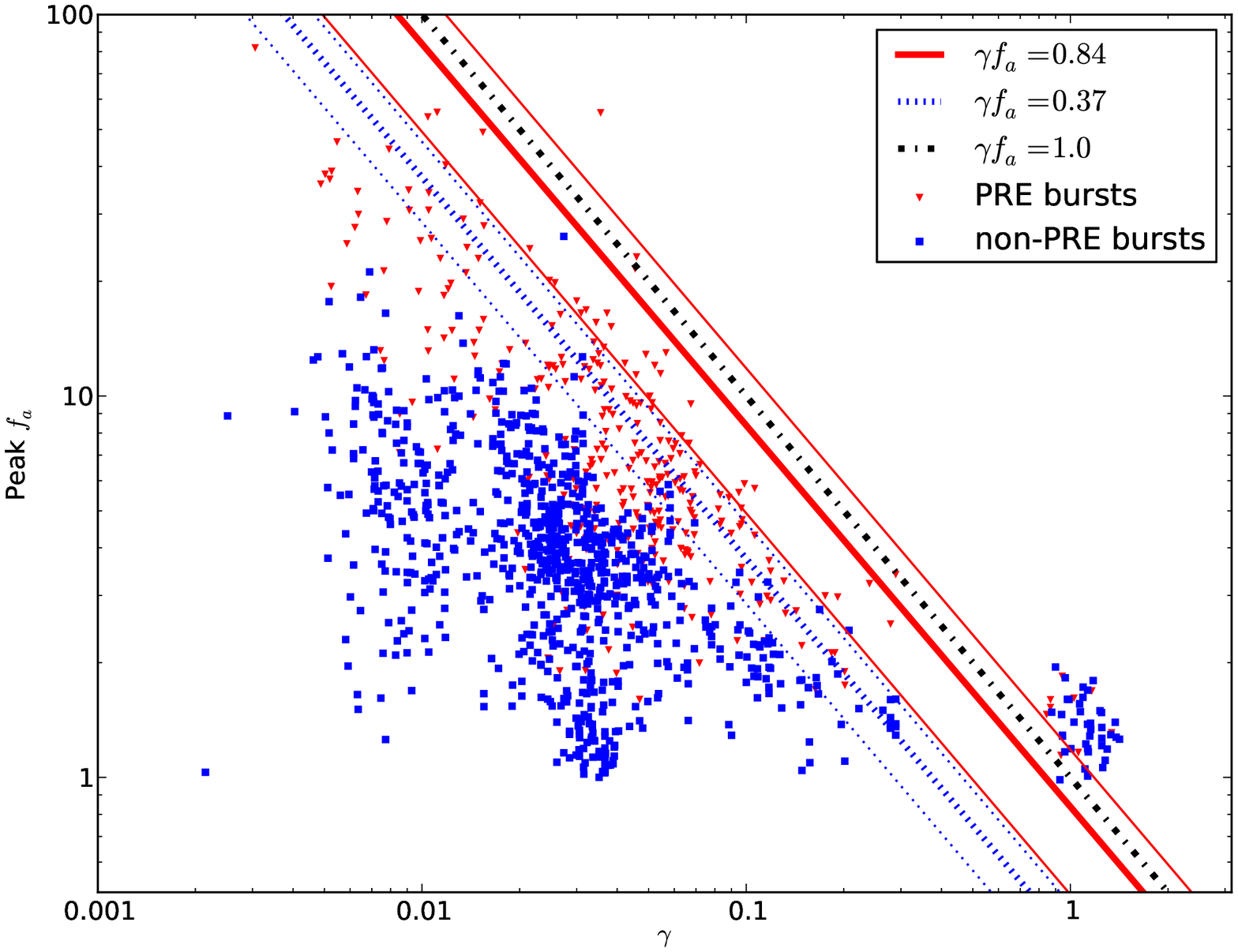}
\caption{Peak $f_a$ against $\gamma$ for every burst in the catalogue from sources for which Eddington fluxes are listed in \citetalias{WorpelEtAl2013}. PRE bursts are shown in red, and non-PRE bursts are shown in blue. There is a clear division between the two classes of bursts. PRE bursts are bounded above by $f_a\gamma\lesssim 0.84$, while the non-PRE bursts are bounded above by $f_a\gamma\lesssim 0.37$ (as calculated by the method explained in \S \ref{sec:Edd_from_nonpre}). The dotted lines indicate $1\sigma$ intervals estimated using the bootstrap method. We also show the hard upper bound $f_a\gamma<1$ (e.g. \citealt{BurgerKatz1983}). The clump of points at $\gamma\approx 1$ represents bursts from the rapid accretors Cyg~X-2 and GX~17+2; these are atypical bursters and we do not consider their bursts in deriving $f_a\gamma$ relations.}
\label{fig:pfa_v_gamma}
\end{figure*}

In Figure \ref{fig:pfa_v_gamma} we show peak $f_a$ against $\gamma$ for every type I burst detected by RXTE. For $0.01<\gamma<0.4$ the non-PRE bursts appear to be bounded above by $f_a \gamma \lesssim (f_a\gamma)_{max}$, where $(f_a\gamma)_{max}$ appears to be somewhat less than 0.5. This would give a lower limit on the Eddington flux roughly twice as constraining as the most conservative estimate $f_a F_{pers}<F_{Edd}$. Determining $\gamma$ requires knowing $F_\text{Edd}$; our method for determining the Eddington fluxes for PRE bursters is given in the Appendix.

We now describe a method by which we estimate the upper bound of $f_a\gamma$ that is reproducible and more objective than estimating by eye. We treat the measured $x_i=(f_a\gamma)_i$ values as normal distributions centered around their measured values, with standard deviations $\sigma_i$ equal to the uncertainty in the $f_a$, times $\gamma$. This normal distribution is given by
\begin{equation}
G(x,\sigma_i,x_i)=\dfrac{1}{\sqrt{2\pi}\sigma_i}\exp\left[-\dfrac{(x-x_i)^2}{2\sigma_i}\right].
\end{equation}

The fraction of this distribution lying above $x_{max}=(f_a\gamma)_{max}$ is
\begin{equation}
\begin{array}{rcl}
   A(\sigma_i,x_i)&=&\displaystyle\int_{x_{max}}^\infty G(x,\sigma_i,x_i)dx\\
   &=&\dfrac{1}{2}\left[1+\erf\left(\dfrac{x_i-x_{max}}{\sqrt{2}\sigma_i}\right)\right], 
\end{array}
\end{equation}

and so the total over all data points is
\begin{equation}
\text{A}_\text{T}=\dfrac{1}{2}\displaystyle\sum_{i=1}^n 1+\erf\left[\dfrac{x_i-x_{max}}{\sqrt{2}\sigma_i}\right],
\end{equation}
where $n$ is the number of data points.

The question is how to find $x_{max}$. First we make a guess $\zeta$ as to the value of $x_{max}$ and calculate an $\text{A}_\text{T}$ from this and the data points $x_i$. Then we take a set of $n$ data points $\hat{x}_i$ with random values between 0 and $\zeta$. We assign to each $\hat{x}_i$ an uncertainty of $\sigma_i$, that is, a random point is given a measured uncertainty. This assumes the real data points are evenly distributed with uncertainties independent of their values, but the procedure can be modified to suit other assumptions. From this random set $\hat{x}_i$ we calculate an expected total area ($\text{A}_\text{T}(rand)$) that can be compared to the real one ($\text{A}_\text{T}(observed)$). If, for instance, $\text{A}_\text{T}(rand) < \text{A}_\text{T}(observed)$ then our guess for $\zeta$ was too high and we try again with a lower $\zeta$. Once the $\zeta$s converge, we take this value to be $x_{max}$. Making many realizations of the random set, we can get an expected $x_{max}$. However, to avoid the analysis assigning too much weight to outliers, we take many bootstrap subsamples of the original data set and take the overall $x_{max}$ to be the average of those from the bootstrap samples; it also gives us uncertainties on $x_{max}$.

\begin{figure*}
\includegraphics[width=68mm]{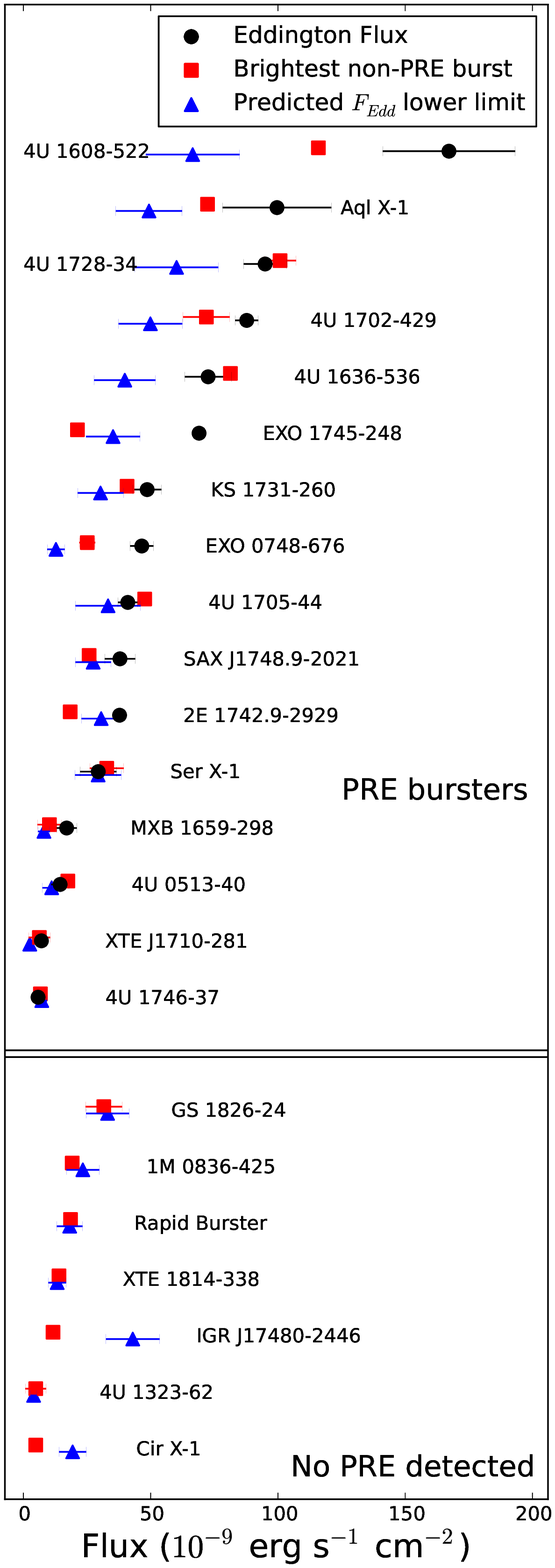}\\
\caption{Lower limits of $F_\text{Edd}$ for various neutron stars from their non-PRE bursts using the method described in \S \ref{sec:Edd_from_nonpre}. We 
have plotted the brightest non-PRE burst (squares) and our new prediction (triangles). For sources with known $F_\text{Edd}$ these are shown (circles) 
to demonstrate that the new method does not overestimate the Eddington flux. For sources where $F_\text{Edd}$ has not been directly measured, we have obtained a 
greatly better constraint for the $F_\text{Edd}$ of Cir X-1 and IGR~17480$-$2446, and a marginally better constraint for 1M~0836$-$425, than 
those given by their brightest non-PRE bursts. For points with no visible $1\sigma$ errorbars, these are smaller than the marker.}
\label{fig:Fedd_prediction}
\end{figure*}

Applying this procedure, we get $x_{max}=0.84\pm0.34$ for the PRE bursts and $x_{max}=0.37\pm0.089$ for the non-PRE bursts. We note that the $f_a\gamma$ relation is a phenomenological one and does not depend on any physical interpretation, though it is consistent with the requirement that the accretion luminosity not exceed Eddington. To test whether our new method gives a more stringent lower bound than the peak burst flux, we compared the two methods for every source for which ten or more non-PRE bursts have been observed, but excluding the rapid accretors Cyg~X$-$2 and GX+17$-$2. We also exclude SLX~1744$-$300, as that source is difficult to distinguish from the nearby ($\Delta\theta\approx 3'$) SLX~1744$-$299 \citep{SkinnerEtAl1990}. We first apply the method to sources for which the Eddington flux is known, to make sure the new method never overestimates the Eddington flux. In 2 of 16 sources, the new method gave a greater lower bound than the brightest non-PRE burst, and in no case did the new method overestimate the Eddington flux beyond observational uncertainties. These results are shown in Figure \ref{fig:Fedd_prediction}.

Applying the new method (with $x_i=(f_aF_\text{Pers})_i$ since $\gamma$ is unknown) to sources with unknown Eddington flux, we get a better bound than the non-PRE bursts for the sources Cir~X$-$1 and IGR~17480$-$2446, with $F_\text{Edd}$ greater than $18.1\pm5.5 \times 10^{-9}$ and $42.0\pm10.5\times 10^{-9}$ erg~s$^{-1}$~cm$^{-2}$ respectively. IGR~17480$-$2446 is located in the globular cluster Terzan 5, with a known distance of about 5.5~kpc (e.g.,\citealt{PapittoEtAl2011}), and a hydrogen accreting neutron star with 1.4$M_\odot$ would have an Eddington flux of 57 $\times 10^{-9}$ erg~s$^{-1}$~cm$^{-2}$. Cir~X$-$1 is believed to be 7.8-10.5~kpc distant \citep{JonkerNelemans2004}. Our result suggests it is probably at the nearer end of that range, since the more distant value would give an Eddington flux of 16 $\times 10^{-9}$ erg~s$^{-1}$~cm$^{-2}$ for a canonical neutron star, slightly higher than our estimate. We also get a marginally more constraining Eddington flux for 1M~0836$-$425 of $23.1\pm6.5 \times 10^{-9}$ erg~s$^{-1}$~cm$^{-2}$.

\section{Conclusion}
\label{sec:conclusion}
We have extended the variable persistent flux method developed by \citet{WorpelEtAl2013} and \citet{intZandEtAl2013} to all type I bursts observed by \RXTE. Our method gives superior spectral fits to type I burst spectra, whether these are radius expansion bursts or not. Our detailed conclusions are as follows:

i) The variable persistent normalization approach developed in \citetalias{WorpelEtAl2013} applies as well to non-PRE bursts as it does to radius expansion bursts. The quality of spectral fits generally improves, and the intensity of the persistent emission usually increases during a burst, typically by a factor of 3-4.

ii) The addition of a third variable parameter increases the uncertainties in the fit parameters. In particular there is a degeneracy between $f_a$ and blackbody burst component normalization, likely due to the indistinctness of the two spectral components.

iii) Photospheric radius expansion adds a confounding spectral effect, likely caused by the changing structure of the photosphere. When this is corrected for, by considering only non-PRE bursts, the spectral fits are much improved, to a Bayes factor of 64. 

iv) Any remaining discrepancy from a blackbody plus variable persistent normalization model is likely to be due to a slight non-blackbody character to the burst emission, rather than spectral shape changes in the persistent emission. This is because changes in the shape of the persistent emission are not detectable in the presence of an additional burst component with the quality of the data currently available. Investigations into the response of the accretion spectrum to a burst transient are desirable, but must await more sensitive instruments such as the \emph{Large Observatory for X-ray Timing} \citep{FerociEtAl2012}.

v) We have found no evidence that the persistent emission spectrum changes shape on timescales of up to an hour before or after a burst, and no major changes in its normalization.

vi) For sources that have never shown a radius expansion burst, we can obtain a lower bound on their Eddington luminosity from the observation that the persistent emission enhancement for non-PRE bursts seldom exceeds a certain fraction (about 37\%) of the Eddington flux. For three sources, this new lower bound was found to be a better constraint than the peak fluxes of the brightest (non-PRE) bursts for those sources.

We recommend that varying the normalization of the presistent flux be adopted as standard for analysing type I bursts, though care must be taken to account for the increased uncertainties in the fit parameters caused by the introduction of a third parameter.
\subsection*{Appendix}
\label{sec:all_Fedd}
\begin{deluxetable*}{lcllc}
\tabletypesize{\scriptsize}
\tablecaption{Supplemental PRE bursts}
\tablewidth{0pt}
\tablehead{
\colhead{Source} & \colhead{Burst Time} & \colhead{Instrument} & \colhead{Flux} & \colhead{Reference} \\
& \colhead{ (MJD) } & & \colhead{ ($10^{-9}\,{\text{erg s}^{-1}\text{ cm}^{-2}}$) } &}
\startdata
GX 17+2 & 44749.0 & Hakucho & 9.3 $\pm$ 1.0 & 1\\
GX 17+2 & 44749.5 & Hakucho & 12.2 $\pm$ 1.1 & 1\\
GX 17+2 & 45194.8 & Hakucho & 10.1 $\pm$ 1.1 & 1\\
GX 17+2 & 45197.6 & Hakucho & 12.8 $\pm$ 1.0 & 1\\
4U 2129+12 & 47454.7 & Ginga & 40.0 & 2\\
SLX 1735-269 & 52897.7 & JEM-X & 57.6 $\pm$ 9.6 & 3\\
SLX 1735-269 & 53108.0 & IBIS/INTEGRAL & 64.0 & 4\\
SLX 1735-269 & 56267.2 & MAXI/GSC & 48.0 $\pm$ 3.0 & 5\\
GX 3+1 & 53248.8 & JEM-X & 53.4 & 6\\
4U 1722-30 & 42666.0 & OSO-8 & 60.0 & 7\\
4U 1746-37 & 46317.0 & EXOSAT & 10.0 $\pm$ 1.0 & 8\\
4U 1746-37 & 46318.0 & EXOSAT & 9.0 $\pm$ 1.0 & 8\\
4U 1820-303 & 42550.0 & SAS-3/RMC & 39.0 $\pm$ 2.0 & 9\\
4U 1820-303 & 42940.0 & SAS-3/RMC & 53.0 $\pm$ 4.2 & 9\\
4U 1820-303 & 42941.0 & SAS-3/RMC & 41.0 $\pm$ 4.5 & 9\\
4U 1820-303 & 42942.0 & SAS-3/RMC & 46.0 $\pm$ 3.7 & 9\\
4U 1820-303 & 42942.0 & SAS-3/RMC & 43.0 $\pm$ 3.9 & 9\\
4U 1820-303 & 42944.0 & SAS-3/RMC & 42.0 $\pm$ 2.9 & 9\\
4U 1820-303 & 46296.0 & EXOSAT & 52.8 $\pm$ 1.9 & 10\\
\enddata
\tablerefs{1. \cite{TawaraEtAl1984}, 2. \cite{vanParadijsEtAl1990}, 3. \cite{MolkovEtAl2005}, 4. \cite{NegoroEtAl2012}, 5. \cite{SgueraEtAl2007}, 6. \cite{ChenevezEtAl2006}, 7. \cite{SwankEtAl1977}, 8. \cite{SztajnoEtAl1987}, 9. \cite{VaccaEtAl1986}, 10. \cite{DamenEtAl1990}}
\label{tab:supplemental_bursts}
\end{deluxetable*}

Table \ref{tab:fedd} lists Eddington fluxes for all sources from which at least one PRE burst has been observed by \RXTE's PCA or WFC instruments. This data is supplemented by a small number of PRE bursts recorded by other X-ray satellites. These are mostly medium to long duration bursts and are listed in Table \ref{tab:supplemental_bursts}. This sample is heterogeneous, but represents our only knowledge of the Eddington flux of the source SLX~1735$-$269, where only non-PRE bursts were observed by \RXTE\ and improves our estimates of the Eddington fluxes of the other sources listed.

The Eddington fluxes for each source are calculated in the same way as in Appendix A of \citetalias{WorpelEtAl2013}. Where no uncertainty on a flux measurement is given in the literature, we have conservatively assumed it is 25\% of the flux measurement. The addition of the supplemental bursts did not significantly change any of the Eddington fluxes listed in \citetalias{WorpelEtAl2013}.

The number of PRE and non-PRE bursts listed in Table \ref{tab:fedd} represent the bursts which made it into our final analysis; bursts discarded for source confusion or the inability to fit a persistent emission model are \emph{not} counted. These are still suitable to estimate the Eddington flux of the source. For example, IGR~17473-2721 has three PRE bursts in the complete \RXTE\ catalog and therefore has a known $F_\text{Edd}$-- but all three of these bursts are source confused so they are not investigated in this work or listed in Table \ref{tab:fedd}.

\section*{Acknowledgements}

H.W. has been supported by an APA postgraduate research scholarship. D.J.P. is the recipient of an Australian Research Council Future Fellowship (project FT130100034). This research utilizes preliminary analysis results from the Multi-INstrument Burst ARchive (MINBAR)\footnote{see \url{ burst.sci.monash.edu/minbar } }, which is supported under the Australian Academy of Science's Scientific Visits to Europe program, and the Australian Research Council's Discovery Projects and Future Fellowship funding schemes. This paper uses results provided by the ASM/RXTE teams at MIT and at the RXTE SOF and GOF at NASA's GSFC. This research made use of Astropy, a community-developed core Python package for Astronomy (Astropy Collaboration, 2013)\nocite{Astropy2013}. This research has made use of data obtained through the High Energy Astrophysics Science Archive Research Center Online Service, provided by the NASA/Goddard Space Flight Center. We are grateful to the International Space Science Institute (ISSI) in Bern for the support of an International Team on Type I X-ray Bursts. We thank the anonymous referee, whose comments significantly improved the paper.

\bibliography{ms}

\begin{thebibliography}{97}
\expandafter\ifx\csname natexlab\endcsname\relax\def\natexlab#1{#1}\fi
\expandafter\ifx\csname href\endcsname\relax
  \def\href#1#2{}\fi
\expandafter\ifx\csname urllinklabel\endcsname\relax
  \def\urllinklabel{[LINK]}\fi
\expandafter\ifx\csname adsurllinklabel\endcsname\relax
  \def\adsurllinklabel{[ADS]}\fi

\bibitem[{{Altamirano} {et~al.}(2008){Altamirano}, {Galloway}, {Chenevez},
  {in't Zand}, {Kuulkers}, {Degenaar}, {Del Monte}, {Feroci}, {Costa},
  {Evangelista}, {Falanga}, {Markwardt}, {Wijnands}, \& {Van Der
  Klis}}]{AltamiranoEtAl2008}
{Altamirano}, D., {Galloway}, D.~K., {Chenevez}, J., {in't Zand}, J.,
  {Kuulkers}, E., {Degenaar}, N., {Del Monte}, E., {Feroci}, M., {Costa}, E.,
  {Evangelista}, Y., {Falanga}, M., {Markwardt}, C., {Wijnands}, R., \& {Van
  Der Klis}, M. 2008, The Astronomer's Telegram, 1651, 1


\bibitem[{{Asai} {et~al.}(2000){Asai}, {Dotani}, {Nagase}, \&
  {Mitsuda}}]{AsaiEtAl2000}
{Asai}, K., {Dotani}, T., {Nagase}, F., \& {Mitsuda}, K. 2000, \apjs, 131, 571


\bibitem[{{Astropy Collaboration} {et~al.}(2013){Astropy Collaboration},
  {Robitaille}, {Tollerud}, {Greenfield}, {Droettboom}, {Bray}, {Aldcroft},
  {Davis}, {Ginsburg}, {Price-Whelan}, {Kerzendorf}, {Conley}, {Crighton},
  {Barbary}, {Muna}, {Ferguson}, {Grollier}, {Parikh}, {Nair}, {Unther},
  {Deil}, {Woillez}, {Conseil}, {Kramer}, {Turner}, {Singer}, {Fox}, {Weaver},
  {Zabalza}, {Edwards}, {Azalee Bostroem}, {Burke}, {Casey}, {Crawford},
  {Dencheva}, {Ely}, {Jenness}, {Labrie}, {Lim}, {Pierfederici}, {Pontzen},
  {Ptak}, {Refsdal}, {Servillat}, \& {Streicher}}]{Astropy2013}
{Astropy Collaboration}, {Robitaille}, T.~P., {Tollerud}, E.~J., {Greenfield},
  P., {Droettboom}, M., {Bray}, E., {Aldcroft}, T., {Davis}, M., {Ginsburg},
  A., {Price-Whelan}, A.~M., {Kerzendorf}, W.~E., {Conley}, A., {Crighton}, N.,
  {Barbary}, K., {Muna}, D., {Ferguson}, H., {Grollier}, F., {Parikh}, M.~M.,
  {Nair}, P.~H., {Unther}, H.~M., {Deil}, C., {Woillez}, J., {Conseil}, S.,
  {Kramer}, R., {Turner}, J.~E.~H., {Singer}, L., {Fox}, R., {Weaver}, B.~A.,
  {Zabalza}, V., {Edwards}, Z.~I., {Azalee Bostroem}, K., {Burke}, D.~J.,
  {Casey}, A.~R., {Crawford}, S.~M., {Dencheva}, N., {Ely}, J., {Jenness}, T.,
  {Labrie}, K., {Lim}, P.~L., {Pierfederici}, F., {Pontzen}, A., {Ptak}, A.,
  {Refsdal}, B., {Servillat}, M., \& {Streicher}, O. 2013, \aap, 558, A33


\bibitem[{{Augusteijn} {et~al.}(1998){Augusteijn}, {van der Hooft}, {de Jong},
  {van Kerkwijk}, \& {van Paradijs}}]{AugusteijnEtAl1998}
{Augusteijn}, T., {van der Hooft}, F., {de Jong}, J.~A., {van Kerkwijk}, M.~H.,
  \& {van Paradijs}, J. 1998, \aap, 332, 561


\bibitem[{{Bagnoli} {et~al.}(2013){Bagnoli}, {in't Zand}, {Galloway}, \&
  {Watts}}]{BagnoliEtAl2013}
{Bagnoli}, T., {in't Zand}, J.~J.~M., {Galloway}, D.~K., \& {Watts}, A.~L.
  2013, \mnras, 431, 1947


\bibitem[{{Barri{\`e}re} {et~al.}(2014){Barri{\`e}re}, {Krivonos}, {Tomsick},
  {Bachetti}, {Boggs}, {Chakrabarty}, {Christensen}, {Craig}, {Hailey},
  {Harrison}, {Hong}, {Mori}, {Stern}, \& {Zhang}}]{BarriereEtAl2014}
{Barri{\`e}re}, N.~M., {Krivonos}, R., {Tomsick}, J.~A., {Bachetti}, M.,
  {Boggs}, S.~E., {Chakrabarty}, D., {Christensen}, F.~E., {Craig}, W.~W.,
  {Hailey}, C.~J., {Harrison}, F.~A., {Hong}, J., {Mori}, K., {Stern}, D., \&
  {Zhang}, W.~W. 2014, ArXiv e-prints


\bibitem[{{Belian} {et~al.}(1976){Belian}, {Conner}, \&
  {Evans}}]{BelianEtAl1976}
{Belian}, R.~D., {Conner}, J.~P., \& {Evans}, W.~D. 1976, \apjl, 206, L135


\bibitem[{{Belloni} {et~al.}(1993){Belloni}, {Hasinger}, {Pietsch},
  {Mereghetti}, {Bignami}, \& {Caraveo}}]{BelloniEtAl1993}
{Belloni}, T., {Hasinger}, G., {Pietsch}, W., {Mereghetti}, S., {Bignami},
  G.~F., \& {Caraveo}, P. 1993, \aap, 271, 487


\bibitem[{{Blissett} \& {Cruise}(1979)}]{BlissettCruise1979}
{Blissett}, R.~J. \& {Cruise}, A.~M. 1979, \mnras, 186, 45


\bibitem[{{Boirin} \& {Parmar}(2003)}]{BoirinParmar2003}
{Boirin}, L. \& {Parmar}, A.~N. 2003, \aap, 407, 1079


\bibitem[{{Burger} \& {Katz}(1983)}]{BurgerKatz1983}
{Burger}, H.~L. \& {Katz}, J.~I. 1983, \apj, 265, 393


\bibitem[{{Cackett} {et~al.}(2006){Cackett}, {Wijnands}, {Linares}, {Miller},
  {Homan}, \& {Lewin}}]{CackettEtAl2006}
{Cackett}, E.~M., {Wijnands}, R., {Linares}, M., {Miller}, J.~M., {Homan}, J.,
  \& {Lewin}, W.~H.~G. 2006, \mnras, 372, 479


\bibitem[{{Campana}(2005)}]{Campana2005}
{Campana}, S. 2005, The Astronomer's Telegram, 535, 1


\bibitem[{{Campana} \& {Stella}(2003)}]{CampanaStella2003}
{Campana}, S. \& {Stella}, L. 2003, \apj, 597, 474


\bibitem[{{Casares} {et~al.}(2006){Casares}, {Cornelisse}, {Steeghs},
  {Charles}, {Hynes}, {O'Brien}, \& {Strohmayer}}]{CasaresEtAl2006}
{Casares}, J., {Cornelisse}, R., {Steeghs}, D., {Charles}, P.~A., {Hynes},
  R.~I., {O'Brien}, K., \& {Strohmayer}, T.~E. 2006, \mnras, 373, 1235


\bibitem[{{Cavecchi} {et~al.}(2011){Cavecchi}, {Patruno}, {Haskell}, {Watts},
  {Levin}, {Linares}, {Altamirano}, {Wijnands}, \& {van der
  Klis}}]{CavecchiEtAl2011}
{Cavecchi}, Y., {Patruno}, A., {Haskell}, B., {Watts}, A.~L., {Levin}, Y.,
  {Linares}, M., {Altamirano}, D., {Wijnands}, R., \& {van der Klis}, M. 2011,
  \apjl, 740, L8


\bibitem[{{Chen} {et~al.}(2013){Chen}, {Zhang}, {Zhang}, {Ji}, {Torres},
  {Kretschmar}, {Li}, \& {Wang}}]{ChenEtAl2013}
{Chen}, Y.-P., {Zhang}, S., {Zhang}, S.-N., {Ji}, L., {Torres}, D.~F.,
  {Kretschmar}, P., {Li}, J., \& {Wang}, J.-M. 2013, \apjl, 777, L9


\bibitem[{{Chenevez} {et~al.}(2006){Chenevez}, {Falanga}, {Brandt},
  {Farinelli}, {Frontera}, {Goldwurm}, {in't Zand}, {Kuulkers}, \&
  {Lund}}]{ChenevezEtAl2006}
{Chenevez}, J., {Falanga}, M., {Brandt}, S., {Farinelli}, R., {Frontera}, F.,
  {Goldwurm}, A., {in't Zand}, J.~J.~M., {Kuulkers}, E., \& {Lund}, N. 2006,
  \aap, 449, L5


\bibitem[{{Church} {et~al.}(1998){Church}, {Parmar}, {Balucinska-Church},
  {Oosterbroek}, {dal Fiume}, \& {Orlandini}}]{ChurchEtAl1998}
{Church}, M.~J., {Parmar}, A.~N., {Balucinska-Church}, M., {Oosterbroek}, T.,
  {dal Fiume}, D., \& {Orlandini}, M. 1998, \aap, 338, 556


\bibitem[{{Church} {et~al.}(2005){Church}, {Reed}, {Dotani},
  {Ba{\l}uci{\'n}ska-Church}, \& {Smale}}]{ChurchEtAl2005}
{Church}, M.~J., {Reed}, D., {Dotani}, T., {Ba{\l}uci{\'n}ska-Church}, M., \&
  {Smale}, A.~P. 2005, \mnras, 359, 1336


\bibitem[{{D'A{\'i}} {et~al.}(2006){D'A{\'i}}, {di Salvo}, {Iaria},
  {M{\'e}ndez}, {Burderi}, {Lavagetto}, {Lewin}, {Robba}, {Stella}, \& {van der
  Klis}}]{DaiEtAl2006}
{D'A{\'i}}, A., {di Salvo}, T., {Iaria}, R., {M{\'e}ndez}, M., {Burderi}, L.,
  {Lavagetto}, G., {Lewin}, W.~H.~G., {Robba}, N.~R., {Stella}, L., \& {van der
  Klis}, M. 2006, \aap, 448, 817


\bibitem[{{Damen} {et~al.}(1990){Damen}, {Magnier}, {Lewin}, {Tan}, {Penninx},
  \& {van Paradijs}}]{DamenEtAl1990}
{Damen}, E., {Magnier}, E., {Lewin}, W.~H.~G., {Tan}, J., {Penninx}, W., \&
  {van Paradijs}, J. 1990, \aap, 237, 103


\bibitem[{{David} {et~al.}(1997){David}, {Goldwurm}, {Murakami}, {Paul},
  {Laurent}, \& {Goldoni}}]{DavidEtAl1997}
{David}, P., {Goldwurm}, A., {Murakami}, T., {Paul}, J., {Laurent}, P., \&
  {Goldoni}, P. 1997, \aap, 322, 229


\bibitem[{{Farinelli} {et~al.}(2007){Farinelli}, {Titarchuk}, \&
  {Frontera}}]{FarinelliEtAl2007}
{Farinelli}, R., {Titarchuk}, L., \& {Frontera}, F. 2007, \apj, 662, 1167


\bibitem[{{Feroci} {et~al.}(2012){Feroci}, {Stella}, {van der Klis},
  {Courvoisier}, {Hernanz}, {Hudec}, {Santangelo}, {Walton}, {Zdziarski}, \&
  et. al.}]{FerociEtAl2012}
{Feroci}, M., {Stella}, L., {van der Klis}, M., {Courvoisier}, T.~J.-L.,
  {Hernanz}, M., {Hudec}, R., {Santangelo}, A., {Walton}, D., {Zdziarski}, A.,
  \& et. al. 2012, Experimental Astronomy, 34, 415


\bibitem[{{Frogel} {et~al.}(1995){Frogel}, {Kuchinski}, \&
  {Tiede}}]{FrogelEtAl1995}
{Frogel}, J.~A., {Kuchinski}, L.~E., \& {Tiede}, G.~P. 1995, \aj, 109, 1154


\bibitem[{{Fujimoto} {et~al.}(1981){Fujimoto}, {Hanawa}, \&
  {Miyaji}}]{FujimotoEtAl1981}
{Fujimoto}, M.~Y., {Hanawa}, T., \& {Miyaji}, S. 1981, \apj, 247, 267


\bibitem[{{Galloway} \& {Lampe}(2012)}]{GallowayLampe2012}
{Galloway}, D.~K. \& {Lampe}, N. 2012, \apj, 747, 75


\bibitem[{{Galloway} {et~al.}(2008){Galloway}, {Muno}, {Hartman}, {Psaltis}, \&
  {Chakrabarty}}]{GallowayEtAl2008}
{Galloway}, D.~K., {Muno}, M.~P., {Hartman}, J.~M., {Psaltis}, D., \&
  {Chakrabarty}, D. 2008, \apjs, 179, 360


\bibitem[{{Galloway} {et~al.}(2006){Galloway}, {Psaltis}, {Muno}, \&
  {Chakrabarty}}]{GallowayEtAl2006}
{Galloway}, D.~K., {Psaltis}, D., {Muno}, M.~P., \& {Chakrabarty}, D. 2006,
  \apj, 639, 1033


\bibitem[{{Gavriil} {et~al.}(2012){Gavriil}, {Strohmayer}, \&
  {Bhattacharyya}}]{GavriilEtAl2012}
{Gavriil}, F.~P., {Strohmayer}, T.~E., \& {Bhattacharyya}, S. 2012, \apj, 753,
  2


\bibitem[{{Gregory}(2005)}]{Gregory2005}
{Gregory}, P.~C. 2005, {Bayesian Logical Data Analysis for the Physical
  Sciences: A Comparative Approach with `Mathematica' Support} (Cambridge
  University Press)


\bibitem[{{Grindlay} {et~al.}(1976){Grindlay}, {Gursky}, {Schnopper},
  {Parsignault}, {Heise}, {Brinkman}, \& {Schrijver}}]{GrindlayEtAl1976}
{Grindlay}, J., {Gursky}, H., {Schnopper}, H., {Parsignault}, D.~R., {Heise},
  J., {Brinkman}, A.~C., \& {Schrijver}, J. 1976, \apjl, 205, L127


\bibitem[{{G{\"u}ver} {et~al.}(2012){G{\"u}ver}, {Psaltis}, \&
  {{\"O}zel}}]{GuverEtAl2012}
{G{\"u}ver}, T., {Psaltis}, D., \& {{\"O}zel}, F. 2012, \apj, 747, 76


\bibitem[{{Hasinger} \& {van der Klis}(1989)}]{HasingervanderKlis1989}
{Hasinger}, G. \& {van der Klis}, M. 1989, \aap, 225, 79


\bibitem[{{Heger} {et~al.}(2007){Heger}, {Cumming}, {Galloway}, \&
  {Woosley}}]{HegerEtAl2007}
{Heger}, A., {Cumming}, A., {Galloway}, D.~K., \& {Woosley}, S.~E. 2007, \apjl,
  671, L141


\bibitem[{{Homan} {et~al.}(2003){Homan}, {Wijnands}, \& {van den
  Berg}}]{HomanEtAl2003}
{Homan}, J., {Wijnands}, R., \& {van den Berg}, M. 2003, \aap, 412, 799


\bibitem[{{Hynes} {et~al.}(2001){Hynes}, {Charles}, {Haswell}, {Casares},
  {Zurita}, \& {Serra-Ricart}}]{HynesEtAl2001}
{Hynes}, R.~I., {Charles}, P.~A., {Haswell}, C.~A., {Casares}, J., {Zurita},
  C., \& {Serra-Ricart}, M. 2001, \mnras, 324, 180


\bibitem[{{Iaria} {et~al.}(2005){Iaria}, {Span{\`o}}, {Di Salvo}, {Robba},
  {Burderi}, {Fender}, {van der Klis}, \& {Frontera}}]{IariaEtAl2005}
{Iaria}, R., {Span{\`o}}, M., {Di Salvo}, T., {Robba}, N.~R., {Burderi}, L.,
  {Fender}, R., {van der Klis}, M., \& {Frontera}, F. 2005, \apj, 619, 503


\bibitem[{{in 't Zand} {et~al.}(1999){in 't Zand}, {Heise}, {Kuulkers},
  {Bazzano}, {Cocchi}, \& {Ubertini}}]{intZandEtAl1999}
{in 't Zand}, J.~J.~M., {Heise}, J., {Kuulkers}, E., {Bazzano}, A., {Cocchi},
  M., \& {Ubertini}, P. 1999, \aap, 347, 891


\bibitem[{{in't Zand} {et~al.}(2013){in't Zand}, {Galloway}, {Marshall},
  {Ballantyne}, {Jonker}, {Paerels}, {Palmer}, {Patruno}, \&
  {Weinberg}}]{intZandEtAl2013}
{in't Zand}, J.~J.~M., {Galloway}, D.~K., {Marshall}, H.~L., {Ballantyne},
  D.~R., {Jonker}, P.~G., {Paerels}, F.~B.~S., {Palmer}, D.~M., {Patruno}, A.,
  \& {Weinberg}, N.~N. 2013, \aap, 553, A83


\bibitem[{{Ji} {et~al.}(2013){Ji}, {Zhang}, {Chen}, {Zhang}, {Torres},
  {Kretschmar}, {Chernyakova}, {Li}, \& {Wang}}]{JiEtAl2013}
{Ji}, L., {Zhang}, S., {Chen}, Y., {Zhang}, S.-N., {Torres}, D.~F.,
  {Kretschmar}, P., {Chernyakova}, M., {Li}, J., \& {Wang}, J.-M. 2013, \mnras,
  432, 2773


\bibitem[{{Ji} {et~al.}(2014{\natexlab{a}}){Ji}, {Zhang}, {Chen}, {Zhang},
  {Torres}, {Kretschmar}, \& {Li}}]{JiEtAl2014}
{Ji}, L., {Zhang}, S., {Chen}, Y., {Zhang}, S.-N., {Torres}, D.~F.,
  {Kretschmar}, P., \& {Li}, J. 2014{\natexlab{a}}, \apj, 782, 40


\bibitem[{{Ji} {et~al.}(2014{\natexlab{b}}){Ji}, {Zhang}, {Chen}, {sZhang},
  {Kretschmar}, {Wang}, \& {Li}}]{JiEtAl2014b}
{Ji}, L., {Zhang}, S., {Chen}, Y.-P., {sZhang}, S.-N., {Kretschmar}, P.,
  {Wang}, J.-M., \& {Li}, J. 2014{\natexlab{b}}, \aap, 564, A20


\bibitem[{{Jonker} {et~al.}(2003){Jonker}, {M{\'e}ndez}, {Nelemans},
  {Wijnands}, \& {van der Klis}}]{JonkerEtAl2003}
{Jonker}, P.~G., {M{\'e}ndez}, M., {Nelemans}, G., {Wijnands}, R., \& {van der
  Klis}, M. 2003, \mnras, 341, 823


\bibitem[{{Jonker} \& {Nelemans}(2004)}]{JonkerNelemans2004}
{Jonker}, P.~G. \& {Nelemans}, G. 2004, \mnras, 354, 355


\bibitem[{{Joss}(1977)}]{Joss1977}
{Joss}, P.~C. 1977, \nat, 270, 310


\bibitem[{{Juett} {et~al.}(2001){Juett}, {Psaltis}, \&
  {Chakrabarty}}]{JuettEtAl2001}
{Juett}, A.~M., {Psaltis}, D., \& {Chakrabarty}, D. 2001, \apjl, 560, L59


\bibitem[{{Keek} {et~al.}(2014){Keek}, {Ballantyne}, {Kuulkers}, \&
  {Strohmayer}}]{KeekEtAl2014}
{Keek}, L., {Ballantyne}, D.~R., {Kuulkers}, E., \& {Strohmayer}, T.~E. 2014,
  ArXiv e-prints


\bibitem[{{Keek} {et~al.}(2008){Keek}, {in't Zand}, {Kuulkers}, {Cumming},
  {Brown}, \& {Suzuki}}]{KeekEtAl2008}
{Keek}, L., {in't Zand}, J.~J.~M., {Kuulkers}, E., {Cumming}, A., {Brown},
  E.~F., \& {Suzuki}, M. 2008, \aap, 479, 177


\bibitem[{{Klein-Wolt} {et~al.}(2007){Klein-Wolt}, {Maitra}, {Wijnands},
  {Swank}, {Markwardt}, \& {Bailyn}}]{Klein-WoltEtAl2007}
{Klein-Wolt}, M., {Maitra}, D., {Wijnands}, R., {Swank}, J.~H., {Markwardt},
  C.~B., \& {Bailyn}, C. 2007, The Astronomer's Telegram, 1070, 1


\bibitem[{{Krimm} {et~al.}(2008){Krimm}, {Kennea}, \&
  {Tueller}}]{KrimmEtAl2008}
{Krimm}, H.~A., {Kennea}, J., \& {Tueller}, J. 2008, The Astronomer's Telegram,
  1432, 1


\bibitem[{{Kuulkers} {et~al.}(2003){Kuulkers}, {den Hartog}, {in't Zand},
  {Verbunt}, {Harris}, \& {Cocchi}}]{KuulkersEtAl2003}
{Kuulkers}, E., {den Hartog}, P.~R., {in't Zand}, J.~J.~M., {Verbunt},
  F.~W.~M., {Harris}, W.~E., \& {Cocchi}, M. 2003, \aap, 399, 663


\bibitem[{{Lattimer} \& {Prakash}(2007)}]{LattimerPrakash2007}
{Lattimer}, J.~M. \& {Prakash}, M. 2007, \physrep, 442, 109


\bibitem[{{Lewin} \& {Joss}(1981)}]{LewinJoss1981}
{Lewin}, W.~H.~G. \& {Joss}, P.~C. 1981, Space Science Reviews, 28, 3


\bibitem[{{Lewin} {et~al.}(1993){Lewin}, {van Paradijs}, \&
  {Taam}}]{LewinEtAl1993}
{Lewin}, W.~H.~G., {van Paradijs}, J., \& {Taam}, R.~E. 1993, \ssr, 62, 223


\bibitem[{{Lin} {et~al.}(2009){Lin}, {Remillard}, \& {Homan}}]{LinEtAl2009}
{Lin}, D., {Remillard}, R.~A., \& {Homan}, J. 2009, \apj, 696, 1257


\bibitem[{{Linares} {et~al.}(2014){Linares}, {Bahramian}, {Heinke}, {Wijnands},
  {Patruno}, {Altamirano}, {Homan}, {Bogdanov}, \& {Pooley}}]{LinaresEtAl2014}
{Linares}, M., {Bahramian}, A., {Heinke}, C., {Wijnands}, R., {Patruno}, A.,
  {Altamirano}, D., {Homan}, J., {Bogdanov}, S., \& {Pooley}, D. 2014, \mnras,
  438, 251


\bibitem[{{Lyu} {et~al.}(2014){Lyu}, {Mendez}, {Sanna}, {Homan}, {Belloni}, \&
  {Hiemstra}}]{LyuEtAl2014}
{Lyu}, M., {Mendez}, M., {Sanna}, A., {Homan}, J., {Belloni}, T., \&
  {Hiemstra}, B. 2014, ArXiv e-prints


\bibitem[{{Maccarone} \& {Coppi}(2003)}]{MaccaroneCoppi2003}
{Maccarone}, T.~J. \& {Coppi}, P.~S. 2003, \aap, 399, 1151


\bibitem[{{Miller} \& {Lamb}(1996)}]{MillerLamb1996}
{Miller}, M.~C. \& {Lamb}, F.~K. 1996, \apj, 470, 1033


\bibitem[{{Molkov} {et~al.}(2005){Molkov}, {Revnivtsev}, {Lutovinov}, \&
  {Sunyaev}}]{MolkovEtAl2005}
{Molkov}, S., {Revnivtsev}, M., {Lutovinov}, A., \& {Sunyaev}, R. 2005, \aap,
  434, 1069


\bibitem[{{Mori} {et~al.}(2005){Mori}, {Maeda}, {Pavlov}, {Sakano}, \&
  {Tsuboi}}]{MoriEtAl2005}
{Mori}, H., {Maeda}, Y., {Pavlov}, G.~G., {Sakano}, M., \& {Tsuboi}, Y. 2005,
  Advances in Space Research, 35, 1137


\bibitem[{{Motta} {et~al.}(2011){Motta}, {D'A{\i}}, {Papitto}, {Riggio}, {di
  Salvo}, {Burderi}, {Belloni}, {Stella}, \& {Iaria}}]{MottaEtAl2011}
{Motta}, S., {D'A{\i}}, A., {Papitto}, A., {Riggio}, A., {di Salvo}, T.,
  {Burderi}, L., {Belloni}, T., {Stella}, L., \& {Iaria}, R. 2011, \mnras, 414,
  1508


\bibitem[{{Muno} {et~al.}(2000){Muno}, {Fox}, {Morgan}, \&
  {Bildsten}}]{MunoEtAl2000}
{Muno}, M.~P., {Fox}, D.~W., {Morgan}, E.~H., \& {Bildsten}, L. 2000, \apj,
  542, 1016


\bibitem[{{Natalucci} {et~al.}(2000){Natalucci}, {Bazzano}, {Cocchi},
  {Ubertini}, {Heise}, {Kuulkers}, {in't Zand}, \& {Smith}}]{NatalucciEtAl2000}
{Natalucci}, L., {Bazzano}, A., {Cocchi}, M., {Ubertini}, P., {Heise}, J.,
  {Kuulkers}, E., {in't Zand}, J.~J.~M., \& {Smith}, M.~J.~S. 2000, \apj, 536,
  891


\bibitem[{{Nath} {et~al.}(2002){Nath}, {Strohmayer}, \& {Swank}}]{NathEtAl2002}
{Nath}, N.~R., {Strohmayer}, T.~E., \& {Swank}, J.~H. 2002, \apj, 564, 353


\bibitem[{{Negoro} {et~al.}(2012){Negoro}, {Asada}, {Serino}, {Nakahira},
  {Morii}, {Ogawa}, {Ueno}, {Tomida}, {Ishikawa}, {Yamaoka}, {Kimura},
  {Mihara}, {Sugizaki}, {Morihana}, {Yamamoto}, {Sugimoto}, {Takagi},
  {Matsuoka}, {Kawai}, {Usui}, {Ishikawa}, {Yoshida}, {Sakamoto}, {Nakano},
  {Tsunemi}, {Sasaki}, {Nakajima}, {Ueda}, {Hiroi}, {Shidatsu}, {Sato},
  {Kawamuro}, {Tsuboi}, {Yamauchi}, {Nishimura}, {Hanayama}, \&
  {Yoshidom}}]{NegoroEtAl2012}
{Negoro}, H., {Asada}, M., {Serino}, M., {Nakahira}, S., {Morii}, M., {Ogawa},
  Y., {Ueno}, S., {Tomida}, H., {Ishikawa}, M., {Yamaoka}, K., {Kimura}, M.,
  {Mihara}, T., {Sugizaki}, M., {Morihana}, K., {Yamamoto}, T., {Sugimoto}, J.,
  {Takagi}, T., {Matsuoka}, M., {Kawai}, N., {Usui}, R., {Ishikawa}, K.,
  {Yoshida}, A., {Sakamoto}, T., {Nakano}, Y., {Tsunemi}, H., {Sasaki}, M.,
  {Nakajima}, M., {Ueda}, Y., {Hiroi}, K., {Shidatsu}, M., {Sato}, R.,
  {Kawamuro}, T., {Tsuboi}, Y., {Yamauchi}, M.~H.~M., {Nishimura}, Y.,
  {Hanayama}, T., \& {Yoshidom}, K. 2012, The Astronomer's Telegram, 4622, 1


\bibitem[{{Oosterbroek} {et~al.}(2001{\natexlab{a}}){Oosterbroek}, {Barret},
  {Guainazzi}, \& {Ford}}]{OosterbroekEtAl2001b}
{Oosterbroek}, T., {Barret}, D., {Guainazzi}, M., \& {Ford}, E.~C.
  2001{\natexlab{a}}, \aap, 366, 138


\bibitem[{{Oosterbroek} {et~al.}(2001{\natexlab{b}}){Oosterbroek}, {Parmar},
  {Sidoli}, {in't Zand}, \& {Heise}}]{OosterbroekEtAl2001}
{Oosterbroek}, T., {Parmar}, A.~N., {Sidoli}, L., {in't Zand}, J.~J.~M., \&
  {Heise}, J. 2001{\natexlab{b}}, \aap, 376, 532


\bibitem[{{Pandel} {et~al.}(2008){Pandel}, {Kaaret}, \&
  {Corbel}}]{PandelEtAl2008}
{Pandel}, D., {Kaaret}, P., \& {Corbel}, S. 2008, \apj, 688, 1288


\bibitem[{{Papitto} {et~al.}(2011){Papitto}, {Bozzo}, {Ferrigno}, {Belloni},
  {Burderi}, {di Salvo}, {Riggio}, {D'A{\`i}}, \& {Iaria}}]{PapittoEtAl2011}
{Papitto}, A., {Bozzo}, E., {Ferrigno}, C., {Belloni}, T., {Burderi}, L., {di
  Salvo}, T., {Riggio}, A., {D'A{\`i}}, A., \& {Iaria}, R. 2011, \aap, 535, L4


\bibitem[{{Papitto} {et~al.}(2010){Papitto}, {Riggio}, {di Salvo}, {Burderi},
  {D'A{\`i}}, {Iaria}, {Bozzo}, \& {Menna}}]{PapittoEtAl2010}
{Papitto}, A., {Riggio}, A., {di Salvo}, T., {Burderi}, L., {D'A{\`i}}, A.,
  {Iaria}, R., {Bozzo}, E., \& {Menna}, M.~T. 2010, \mnras, 407, 2575


\bibitem[{{Peille} {et~al.}(2014){Peille}, {Olive}, \&
  {Barret}}]{PeilleEtAl2014}
{Peille}, P., {Olive}, J., \& {Barret}, D. 2014, ArXiv e-prints


\bibitem[{{Piraino} {et~al.}(2007){Piraino}, {Santangelo}, {di Salvo},
  {Kaaret}, {Horns}, {Iaria}, \& {Burderi}}]{PirainoEtAl2007}
{Piraino}, S., {Santangelo}, A., {di Salvo}, T., {Kaaret}, P., {Horns}, D.,
  {Iaria}, R., \& {Burderi}, L. 2007, \aap, 471, L17


\bibitem[{{Sakano} {et~al.}(2002){Sakano}, {Koyama}, {Murakami}, {Maeda}, \&
  {Yamauchi}}]{SakanoEtAl2002}
{Sakano}, M., {Koyama}, K., {Murakami}, H., {Maeda}, Y., \& {Yamauchi}, S.
  2002, \apjs, 138, 19


\bibitem[{{Sala} {et~al.}(2012){Sala}, {Haberl}, {Jos{\'e}}, {Parikh},
  {Longland}, {Pardo}, \& {Andersen}}]{SalaEtAl2012}
{Sala}, G., {Haberl}, F., {Jos{\'e}}, J., {Parikh}, A., {Longland}, R.,
  {Pardo}, L.~C., \& {Andersen}, M. 2012, \apj, 752, 158


\bibitem[{{Sguera} {et~al.}(2007){Sguera}, {Bazzano}, \&
  {Bird}}]{SgueraEtAl2007}
{Sguera}, V., {Bazzano}, A., \& {Bird}, A.~J. 2007, The Astronomer's Telegram,
  1340, 1


\bibitem[{{Sidoli} {et~al.}(2005){Sidoli}, {Parmar}, \&
  {Oosterbroek}}]{SidoliEtAl2005}
{Sidoli}, L., {Parmar}, A.~N., \& {Oosterbroek}, T. 2005, \aap, 429, 291


\bibitem[{{Skinner} {et~al.}(1990){Skinner}, {Foster}, {Willmore}, \&
  {Eyles}}]{SkinnerEtAl1990}
{Skinner}, G.~K., {Foster}, A.~J., {Willmore}, A.~P., \& {Eyles}, C.~J. 1990,
  \mnras, 243, 72


\bibitem[{{Stella} \& {Rosner}(1984)}]{StellaRosner1984}
{Stella}, L. \& {Rosner}, R. 1984, \apj, 277, 312


\bibitem[{{Strohmayer} \& {Bildsten}(2006)}]{StrohmayerBildsten2006}
{Strohmayer}, T. \& {Bildsten}, L. 2006, in {Compact Stellar X-ray Sources},
  ed. {{Lewin}, W.~H.~G.~and~{van der Klis}, M.}, 113--156


\bibitem[{{Strohmayer} \& {Brown}(2002)}]{StrohmayerBrown2002}
{Strohmayer}, T.~E. \& {Brown}, E.~F. 2002, \apj, 566, 1045


\bibitem[{{Swank} {et~al.}(1977){Swank}, {Becker}, {Boldt}, {Holt}, {Pravdo},
  \& {Serlemitsos}}]{SwankEtAl1977}
{Swank}, J.~H., {Becker}, R.~H., {Boldt}, E.~A., {Holt}, S.~S., {Pravdo},
  S.~H., \& {Serlemitsos}, P.~J. 1977, \apjl, 212, L73


\bibitem[{{Sztajno} {et~al.}(1987){Sztajno}, {Fujimoto}, {van Paradijs},
  {Vacca}, {Lewin}, {Penninx}, \& {Trumper}}]{SztajnoEtAl1987}
{Sztajno}, M., {Fujimoto}, M.~Y., {van Paradijs}, J., {Vacca}, W.~D., {Lewin},
  W.~H.~G., {Penninx}, W., \& {Trumper}, J. 1987, MNRAS, 226, 39


\bibitem[{{Tawara} {et~al.}(1984){Tawara}, {Hirano}, {Kii}, {Matsuoka}, \&
  {Murakami}}]{TawaraEtAl1984}
{Tawara}, Y., {Hirano}, T., {Kii}, T., {Matsuoka}, M., \& {Murakami}, T. 1984,
  \pasj, 36, 861


\bibitem[{{Thompson} {et~al.}(2005){Thompson}, {Rothschild}, {Tomsick}, \&
  {Marshall}}]{ThompsonEtAl2005}
{Thompson}, T.~W.~J., {Rothschild}, R.~E., {Tomsick}, J.~A., \& {Marshall},
  H.~L. 2005, \apj, 634, 1261


\bibitem[{{Titarchuk}(1994)}]{Titarchuk1994b}
{Titarchuk}, L. 1994, \apj, 434, 570


\bibitem[{{Vacca} {et~al.}(1986){Vacca}, {Lewin}, \& {van
  Paradijs}}]{VaccaEtAl1986}
{Vacca}, W.~D., {Lewin}, W.~H.~G., \& {van Paradijs}, J. 1986, \mnras, 220, 339


\bibitem[{{van Paradijs} {et~al.}(1990){van Paradijs}, {Dotani}, {Tanaka}, \&
  {Tsuru}}]{vanParadijsEtAl1990}
{van Paradijs}, J., {Dotani}, T., {Tanaka}, Y., \& {Tsuru}, T. 1990, \pasj, 42,
  633


\bibitem[{{van Paradijs} \& {Lewin}(1986)}]{vanParadijsLewin1986}
{van Paradijs}, J. \& {Lewin}, H.~G. 1986, \aap, 157, L10


\bibitem[{{Walker}(1992)}]{Walker1992}
{Walker}, M.~A. 1992, ApJ, 385, 642


\bibitem[{{Wang} {et~al.}(2001){Wang}, {Chakrabarty}, {Roche}, {Charles},
  {Kuulkers}, {Shahbaz}, {Simpson}, {Forbes}, \& {Helsdon}}]{WangEtAl2001}
{Wang}, Z., {Chakrabarty}, D., {Roche}, P., {Charles}, P.~A., {Kuulkers}, E.,
  {Shahbaz}, T., {Simpson}, C., {Forbes}, D.~A., \& {Helsdon}, S.~F. 2001,
  \apjl, 563, L61


\bibitem[{{Werner} {et~al.}(2004){Werner}, {in't Zand}, {Natalucci},
  {Markwardt}, {Cornelisse}, {Bazzano}, {Cocchi}, {Heise}, \&
  {Ubertini}}]{WernerEtAl2004}
{Werner}, N., {in't Zand}, J.~J.~M., {Natalucci}, L., {Markwardt}, C.~B.,
  {Cornelisse}, R., {Bazzano}, A., {Cocchi}, M., {Heise}, J., \& {Ubertini}, P.
  2004, \aap, 416, 311


\bibitem[{{Woosley} \& {Taam}(1976)}]{WoosleyTaam1976}
{Woosley}, S.~E. \& {Taam}, R.~E. 1976, \nat, 263, 101


\bibitem[{{Worpel} {et~al.}(2013){Worpel}, {Galloway}, \&
  {Price}}]{WorpelEtAl2013}
{Worpel}, H., {Galloway}, D.~K., \& {Price}, D.~J. 2013, \apj, 772, 94


\bibitem[{{Younes} {et~al.}(2009){Younes}, {Boirin}, \&
  {Sabra}}]{YounesEtAl2009}
{Younes}, G., {Boirin}, L., \& {Sabra}, B. 2009, \aap, 502, 905


\end{thebibliography}

\end{document}